\documentclass[sigconf]{acmart} 
\usepackage{xcolor} 
\usepackage{multirow}
\usepackage{booktabs} 
\usepackage{amsmath}
\usepackage{graphicx}
\usepackage[belowskip=-12pt]{caption}
\usepackage{subcaption}
\usepackage{rotating}
\usepackage{tabularx}
\usepackage{wrapfig}
\usepackage{listings}
\usepackage{color, colortbl}
\usepackage{balance} 
\usepackage{lscape}

\usepackage{hyperref}
\usepackage{xspace}
\usepackage{framed}
\usepackage{syntax}
\usepackage[framemethod=TikZ]{mdframed}
\usepackage{soul}
\usepackage{flushend}
\usepackage{enumitem}
   
\usepackage{tikz}
\usepackage[most]{tcolorbox}

\newcommand*\redcheckbox{\tikz \fill[red!70!blue] [rounded corners=1.2] (0,0) (0,.27) (0,0) (0, 0) rectangle ++(.2,.2);}
\newcommand*\bluecheckbox{\tikz \fill[blue!60!yellow] [rounded corners=1.2] (0,0) (0,.27) (0,0) (0, 0) rectangle ++(.2,.2);}
\newcommand*\greencheckbox{\tikz \fill[green!60!blue] [rounded corners=1.2] (0,0) (0,.27) (0,0) (0, 0) rectangle ++(.2,.2);}
\newcommand*\yellowcheckbox{\tikz \fill[yellow!60!red] [rounded corners=1.2] (0,0) (0,.27) (0,0) (0, 0) rectangle ++(.2,.2);}

\definecolor{codegreen}{rgb}{0,0.6,0}
\definecolor{codegray}{rgb}{0.5,0.5,0.5}
\definecolor{codepurple}{rgb}{0.58,0,0.82}
\definecolor{backcolour}{rgb}{0.95,0.95,0.92}
\definecolor{Gray}{gray}{0.1}

\lstdefinestyle{mystyle}{
	backgroundcolor=\color{backcolour},   
	commentstyle=\color{codegreen},
	keywordstyle=\color{magenta},
	numberstyle=\tiny\color{codegray},
	stringstyle=\color{codepurple},
	basicstyle=\scriptsize,
	breakatwhitespace=false,         
	breaklines=true,                 
	captionpos=b,                    
	keepspaces=true,                 
	numbers=left,                    
	numbersep=5pt,                  
	showspaces=false,                
	showstringspaces=false,
	showtabs=false,                  
	tabsize=2,
	xleftmargin=.015\textwidth, 
	otherkeywords={self}
}

\lstdefinelanguage{Pythonna}{%
	language     = python,
	morekeywords = {to_categorical, flow_from_directory, pad_sequences, load_image}
}

\lstdefinestyle{customc}{
	belowcaptionskip=1\baselineskip,
	breaklines=false,
	frame= single,
	breaklines = true,
	xleftmargin=\parindent,
	language= Pythonna,
	showstringspaces=false,
	basicstyle=\footnotesize\ttfamily,
	keywordstyle=\bfseries\color{green!40!black},
	commentstyle=\itshape\color{purple!40!black},
	identifierstyle=\color{blue},
	stringstyle=\color{codegreen},
	backgroundcolor=\color{gray!4}
}

\graphicspath{{./figures/}}

\newcommand{\tabref}[1]{Table~\ref{#1}}
\newcommand{\fignref}[1]{Figure~\ref{#1}}

\newcommand{\secref}[1]{\S\ref{#1}}
\newcommand{\secnref}[1]{\S\ref{#1}}

\newcommand{\sof}{\textit{Stack Overflow}\xspace}

\newcommand{\gh}{GitHub\xspace}

\newcounter{rqs}
\stepcounter{rqs}

\newcounter{NumObservations}
\stepcounter{NumObservations}
\definecolor{shadecolor}{rgb}{.9,.9,.9}

\newcommand{\findings}[1]{%
\begin{mdframed}[
	backgroundcolor= yellow!10,
	linewidth=.2pt,
	linecolor = gray!110,
	roundcorner=2pt,
	skipabove=6pt,
	skipbelow=-1pt
	innertopmargin=7pt,
	innerbottommargin=2pt,
	innerrightmargin=4pt,
	innerleftmargin=2pt,
	leftmargin = 0pt,
	rightmargin = 0pt]%
	\noindent{\textbf{Finding \arabic{NumObservations}}: \em #1} 
\end{mdframed}%
	\stepcounter{NumObservations}
}

\newtcbtheorem{Summary}{\bfseries Summary}{enhanced,drop shadow={black!50!white},
  coltitle=black,
  top=0.3in,
  attach boxed title to top left=
  {xshift=1.5em,yshift=-\tcboxedtitleheight/2},
  boxed title style={size=small,colback=pink}
}{summary}

\newtcolorbox[auto counter]{summary}[1][]{title={\bfseries Summary~\thetcbcounter},enhanced,drop shadow={black!50!white},
  coltitle=black,
  top=0.3in,
  attach boxed title to top left=
  {xshift=1.5em,yshift=-\tcboxedtitleheight/2},
  boxed title style={size=small,colback=pink},#1}

\mdfdefinestyle{mpdframe}{
    frametitlebackgroundcolor   =black!15,
    frametitlerule              =true,
    roundcorner                 =4pt,
    middlelinewidth             =0.5pt,
    innermargin                 =0cm,
    outermargin                 =0.0cm,
    innerleftmargin             =0.15cm,
    innerrightmargin            =0.15cm,
    innertopmargin              =0.15cm,
    innerbottommargin           =0.15cm
}

\AtBeginDocument{%
  \providecommand\BibTeX{{%
    \normalfont B\kern-0.5em{\scshape i\kern-0.25em b}\kern-0.8em\TeX}}}

\copyrightyear{2022}
\acmYear{2022}
\setcopyright{rightsretained}
\acmConference[ICSE '22]{44th International Conference on Software Engineering}{May 21--29, 2022}{Pittsburgh, PA, USA}
\acmBooktitle{44th International Conference on Software Engineering (ICSE '22), May 21--29, 2022, Pittsburgh, PA, USA}
\acmDOI{10.1145/3510003.3510057}
\acmISBN{978-1-4503-9221-1/22/05}

\begin{document}

\title{Data Science Lifecycles: A Comprehensive Study of Definitions and Frameworks}
\title{Data Science Lifecycles: A Study of Software and Processes}
\title{What is a Data Science Lifecycle? A View From Above and Below}
\title{Data Science Pipelines: A Comprehensive Study of Patterns}
\title{Data Science Pipelines: A Comprehensive Study of Art \& Practice}
\title{The Art and Practice of Data Science Pipelines}
\subtitle{A Comprehensive Study of Data Science Pipelines In Theory, In-The-Small, and In-The-Large}

\author{Sumon Biswas}
\affiliation{
	\institution{Iowa State University}
	\city{Ames}
	\state{IA}
	\country{USA}}
\email{sumon@iastate.edu}

\author{Mohammad Wardat}
\affiliation{%
	\institution{Iowa State University}
	\city{Ames}
	\state{IA}
	\country{USA}}
\email{wardat@iastate.edu}

\author{Hridesh Rajan}
\affiliation{
	\institution{Iowa State University}
	\city{Ames}
	\state{IA}
	\country{USA}}
\email{hridesh@iastate.edu}

\begin{abstract}

Increasingly larger number of software systems today are including data science components for descriptive, predictive, and prescriptive analytics. The collection of data science stages from acquisition, to cleaning/curation, to modeling, and so on are referred to as {\em data science pipelines}. To facilitate research and practice on data science pipelines, it is essential to understand their nature. What are the typical stages of a data science pipeline? How are they connected? Do the pipelines differ in the theoretical representations and that in the practice? 
Today we do not fully understand these architectural characteristics of data science pipelines. In this work, we present a three-pronged comprehensive study to answer this for the state-of-the-art, data science in-the-small, and data science in-the-large. Our study analyzes three datasets: a collection of 71 proposals for data science pipelines and related concepts in {\em theory}, a collection of over 105 implementations of curated data science pipelines from Kaggle competitions to understand data science {\em in-the-small}, and a collection of 21 mature data science projects from \gh to understand data science {\em in-the-large}. Our study has led to three representations of data science pipelines that capture the essence of our subjects in theory, in-the-small, and in-the-large.

\end{abstract}


   
 
\maketitle

\section{Introduction}
\label{sec:intro}

Data science processes, also called {\em data science stages} as in stages 
of a pipeline, for descriptive, predictive, and prescriptive 
analytics are becoming integral components of many software systems today.
The data science stages are organized into a {\em data science pipeline}, 
where data might flow from one stage in the pipeline to the next.
These data science stages generally perform different tasks such as 
data acquisition, data preparation, storage, feature engineering, 
modeling, training, evaluation of the machine learning model, etc. 
In order to design and build software systems with data science stages 
effectively, we must understand the structure of the data science pipelines. 
Previous work 
has shown that understanding the structure and patterns used in existing systems 
and literature can help build better systems ~\cite{SoftwareArchitecture,DesignPatterns}.
In this work, we have taken the first step to understand the structure and patterns of DS pipelines. 

Fortunately, we have a number of instances in both the state-of-the-art and practice to draw observations.
In the literature, there have been a number of proposals to organize data
science pipelines. We call such proposals {\em DS Pipelines in theory}.
Another source of information is Kaggle, a widely known platform for 
data scientists to host and participate in DS competitions, 
share datasets, machine learning models, and code.
Kaggle contains a large number of data science pipelines, but these 
pipelines are typically developed by a single data scientist as small standalone programs.
We call such instances {\em DS Pipelines in-the-small}.
The third source of DS pipelines are mature data science projects 
on \gh developed by teams, suitable for reuse. 
We call such instances {\em DS Pipelines in-the-large}.

This work presents a study of DS pipelines in theory, in-the-small, 
and in-the-large. 
We studied 71 different proposals for DS pipelines and related concepts from the literature. 
We also studied 105 instances of DS pipelines from Kaggle.
Finally, we studied 21 matured open-source data science projects
from \gh. 
For both Kaggle and \gh , we selected projects that make use of Python to ease comparative analysis. 
In each setting, we answer the following overarching questions.

\begin{enumerate}[leftmargin=*, topsep=2.5pt] 
	\item \textbf{Representative pipeline:}
	What are the stages in DS pipeline and how frequently they appear?
	\item \textbf{Organization:}
	How are the pipeline stages organized?
	\item \textbf{Characteristics:}
	What are the characteristics of the pipelines in a setting and how does that compare with the others?
\end{enumerate}

This work attempts to inform the terminology and practice for 
designing DS pipeline.
We found that DS pipelines differ significantly in terms of detailed 
structures and patterns among theory, in-the-small, and in-the-large. 
Specifically, a number of stages are absent in-the-small, and the pipelines have a more linear structure with an emphasis on data exploration.
Out of the eleven stages seen in theory, only six stages are present 
in pipeline in-the-small, 
namely \textit{data collection}, \textit{data preparation}, 
\textit{modeling}, \textit{training}, \textit{evaluation}, and \textit{prediction}. 
In addition, pipelines in-the-small do not have clear separation between stages
which makes the maintenance harder.
On the other hand, the DS pipelines in-the-large have a more complex
structure with feedback loops and sub-pipelines.
We identified different pipeline patterns followed in specific phase (development/post-development) of the large DS projects. 
The abstraction of stages are stricter in-the-large having both loosely- and tightly-coupled structure.

Our investigation also suggest that DS pipeline is a well used software architecture but often built in ad hoc manner. We demonstrated the importance of standardization and analysis framework for DS pipeline following the traditional software engineering research on software architecture and design patterns \cite{SoftwareArchitecture, lientz1978characteristics, parnas1985modular}.
We contributed three representations of DS pipelines that 
capture the essence of our subjects in theory, in-the-small, and in-the-large that would facilitate building new DS systems. 
We anticipate our results to inform design decisions made by the pipeline architects, 
practitioners, and software engineering teams.
Our results will also help the DS researchers and developers to 
identify whether the pipeline is missing any important stage or feedback 
loops (e.g., \textit{storage} and \textit{evaluation} are missed in many pipelines).

The rest of this paper is organized as follows: 
in section \secref{sec:theory}, we present our study of DS pipelines in theory. 
Section \secref{sec:in-the-small} describes our study of DS pipelines in-the-small.
In section \secref{sec:in-the-large}, we describe our study of DS pipelines in-the-large.
Section \secref{sec:discussion} discusses the implications,
section \secref{sec:threat} describes the threats to the validity,
section \secref{sec:related} describes related work, 
and section \secref{sec:conc} concludes.

\section{DS Pipeline in Theory}
\label{sec:theory}

\textbf{Data Science}. 
Data Science (DS) is a broad area that brings together computational
understanding, inferential thinking, and the knowledge of the application area. 
Wing \cite{wing2019data} argues that DS studies how to extract value out of data.
However, the value of data and extraction process depends on the application and context. 
DS includes a broad set of traditional disciplines such as data management, 
data infrastructure building, data-intensive algorithm development, 
AI (machine learning and deep learning), etc., 
that covers both the fundamental and practical perspectives
from computer science, mathematics, statistics, and domain-specific knowledge~\cite{berman2018realizing, todd2017computing}. 
DS also incorporates the business, organization, policy and privacy issues 
of data and data-related processes. 
Any DS project involves three main stages: data collection and
preparation, analysis and modeling, and finally deployment~\cite{wickham2019data}. 
DS is also more than statistics or data mining since it incorporates understanding 
of data and its pattern, developing important questions and answering them, 
and communicating results~\cite{todd2017computing}.

\textbf{Data Science Pipeline}. The term 
\textit{pipeline} was introduced by \citeauthor{garlan2000software} with 
\textit{box-and-line} diagrams and explanatory prose that assist software developers 
to design and describe complex systems so that the software becomes intelligible~\cite{garlan2000software}. 
\citeauthor{SoftwareArchitecture} have provided the \textit{pipes-and-filter} 
design pattern that involves stages with processing units (filters) and 
ordered connections (pipes)~\cite{SoftwareArchitecture}. 
They also argued that pipeline gives proper semantics and vocabulary which 
helps to describe the concerns, constraints, relationship between the 
sub-systems, and overall computational paradigm~\cite{garlan2000software, SoftwareArchitecture}. 
By {\em data science pipeline} (DS pipeline), we are referring to a series 
of processing {\em stage}s that interact with data, usually acquisition, 
management, analysis, and reasoning~\cite{olson2016evaluation, nguyen2019machine}. 
The sequential DS stages from acquisition, to cleaning/curation, to modeling, 
and so on are referred to as {\em data science pipeline}. 
A DS pipeline may consist of several stages and connections between them. 
The stages are defined to perform particular tasks and connected to other stage(s) 
with input-output relations~\cite{msml}. 
However, the definitions of the stages are not consistent across studies 
in the literature. The terminology vary depending on the application context and focus.

Different study in the literature presented DS pipeline based on their context and desiderata. 
No study has been conducted to unify the notions DS pipeline and collect the 
concepts~\cite{sculley2015hidden}. 
While designing a new DS pipeline~\cite{wirth2000crisp}, dividing roles in 
DS teams~\cite{kim2016emerging}, defining software process in data-intensive 
setting~\cite{wan2019does}, identifying best practices in AI and modularizing 
DS components~\cite{msml}, it is important to understand the current state of 
the DS pipeline, its variations and different stages. 
To understand the DS pipelines and compare them, we collected the available 
pipelines from the literature and conducted an empirical study to unify the 
stages with their subtasks. 
Then we created a representative DS pipeline with the definitions of the stages.
Next, we present the methodology and results of our 
analysis of DS pipelines in theory.

\subsection{Methodology}

\subsubsection{Collecting Data Science Pipelines}

We searched for the studies published in the literature and popular press 
that describes DS pipelines. 
We considered the studies that described both end-to-end DS pipeline or 
a partial DS pipeline specific to a context.
First, we searched for peer-reviewed papers 
published in the last decade i.e., from 2010 to 2020. 
We searched the terms ``\textit{data science pipeline}'', 
``\textit{machine learning pipeline}'', ``\textit{big data lifecycle}'',
``\textit{deep learning workflow}'', and the permutation of these 
keywords in IEEE Xplore, ACM Digital Library and Google Scholar. 
From a large pool, we selected 1,566 papers that fall broadly in 
the area of computer science, software engineering and data science. 
Then we analyzed each article in this pool to select the ones that 
propose or describe a DS pipeline. 
We found many papers in this collection use the terms (e.g., 
ML lifecycle), but do not contain a DS pipeline. 
We selected the ones that contain DS pipeline and extracted the 
pipelines (screenshot/description) as evidence from the article. 
The extracted raw pipelines are available in the artifact accompanied 
by this paper~\cite{sumon2021pipeline}. 
Thus, we found 46 DS pipelines that were published in the last decade.

Besides peer-reviewed papers, by searching the keywords on web, 
we collected the DS pipelines from US patent, industry blogs 
(e.g., Microsoft, GoogleCloud, IBM blogs), and popular press published 
between 2010 and 2020.
After manual inspection, we found 25 DS pipelines from this grey literature. 
Thus, we collected 71 subjects (46 from peer reviewed articles 
and 25 from grey literature) that contain DS pipeline. 
We used an open-coding method to analyze these DS pipelines in theory~\cite{sumon2021pipeline} .

\subsubsection{Labeling Data Science Pipelines}
\label{ldsp}

In the collected references, DS pipeline is defined with a set of stages
(\textit{data acquisition}, \textit{data preparation}, \textit{modeling}, etc.) and connections among them.
Each stage in the pipeline is defined for performing a specific task
and connected to other stages. However, not all the
studies depict DS pipelines with the same set of stages and connections. 
The studies use different terminologies for defining the stages depending on the context. 
To be able to compare the pipelines, we had to understand the definitions 
and transform them into a canonical form.
For a given DS pipeline, identifying their stages and mapping them to a
canonical form is often challenging.
The sub-tasks, overall
goal of the project, utilities affect the understanding of the pipeline stages. 
To counter these challenges, we used an open-coding method to label the stages of the pipelines. 

\begin{figure}[t]
	\centering
	\includegraphics[width=0.75\columnwidth]{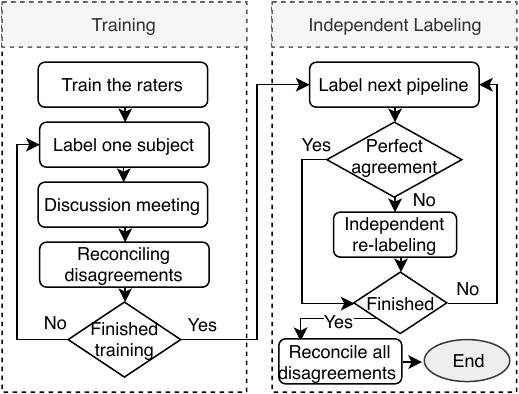}
	\caption{Labeling method for DS pipelines in theory}
	\label{label}
\end{figure}

Two authors labeled the collected DS pipelines into different criteria. 
Each author read the article, understood the pipeline, identified the stages, 
and labeled them. 
In each iteration, the raters labeled 10\% of the subjects (7-8 pipelines). 
The first 8 subjects were used for training and forming the initial labels. 
After each iteration, we calculated the Cohen's Kappa coefficient~\cite{viera2005understanding}, 
identified the mismatches, and resolved them in the presence of a moderator, 
who is another author.
Thus, we found the representative DS pipeline after rigorous discussions 
among the raters and the moderator. 
The methodology of this open-coding procedure is shown in \fignref{label}. 
The entire labeling process was divided into two phases: 1) training, 
and 2) independent labeling.

\textbf{Training:}
The two raters were trained on the goal of this project and their roles. 
We randomly selected eight subjects for training. 
First, the raters and the moderator had discussions on three subjects 
and identified the stages in their DS pipeline.
Thus, we formed the commonly occurred stages and their definitions, 
which were updated through the entire labeling and reconciliation process later.
After the initial discussion and training, the raters were given the 
already created definitions of the stages and one pipeline from 
the remaining five for training.
The raters labeled this pipeline independently. 
After labeling the pipeline, we calculated the agreement and 
conducted a discussion session among the raters and the moderator. 
In this session, we reconciled the disagreements and updated the 
labels with the definitions. 
We continued the training session until we got perfect agreement independently.
The inter-rater agreement was calculated using Cohen's Kappa coefficient \cite{viera2005understanding}. A higher $\kappa$ ([0, 1]) indicates a better agreement. 
The interpretation of of $\kappa$ is shown in \fignref{tab:kappa-table}.
In the discussion meetings, the raters discussed each label
(both agreed and disagreed ones) with the other rater and moderator, argued for the disagreed ones and reconciled them. In this way, we came up with most of the stages and a representative terminology for each stage including the sub-tasks.

\textbf{Independent labeling:} 
After completing the training session, the rest of the subjects were labeled independently by the raters. 
The raters labeled the remaining 63 labels: 7 subjects (10\%) in each of the 9 iterations. The distribution of $\kappa$ after each independent labeling iteration is shown in \fignref{tab:kappa-calc}. In each iteration, first, the raters had the labeling session, and then the raters and moderator had the reconciliation session. 

\textit{Labeling.}
The raters labeled separately so that their labels were private, and they did not discuss while labeling.
The raters identified the stages and connections between them, and finally labeled whether the DS pipeline involves processes related to cyber, physical or human component in it. 
In independent labeling, we found almost perfect agreement ($\kappa$
= 0.83) on average. Even after high agreement, there were very few
disagreements in the labels, which were reconciled after each iteration.

\textit{Reconciling.} Reconciliation happened for each label for the subject studies in the training session, and the disagreed labels for the studies in independent labeling session.
In training session, the reconciliation was done in discussion meetings among the raters and the moderator, whereas for the independent labels, reconciliation was done by the moderator after separate meetings with the two raters. For reconciliation, the raters described their arguments for the mislabeled stages. 
For a few cases, we had straightforward solution to go for one label. For others, both the raters had good arguments for their labels, and we had to decide on that label by updating the stages in the definition of the pipeline. All the labeled pipelines from the subjects are shared in our paper artifact~\cite{sumon2021pipeline}.

\begin{figure}[t]
	\footnotesize
	\setlength\tabcolsep{3pt}
	\begin{subtable}{.5\linewidth}
		\centering
		  \begin{tabular}{|l|l|}
			\hline
			\textbf{Range ($\kappa$)}       & \textbf{Agreement level}       \\ \hline
			0.00 - 0.20 & Slight agreement      \\ \hline
			0.21 - 0.40 & Fair agreement        \\ \hline
			0.41 - 0.60 & Moderate agreement    \\ \hline
			0.61 - 0.80 & Substantial agreement \\ \hline
			0.81 - 1.00 & Perfect agreement     \\ \hline
		\end{tabular}
		  \caption{Interpretation of Kappa ($\kappa$)}
		  \label{tab:kappa-table}
	\end{subtable}%
	\begin{subtable}{.5\linewidth}
		\centering
		  \begin{tabular}{|l|l||l|l|}
			\hline
			\textbf{Iteration \#}       & \textbf{$\kappa$} & \textbf{Iteration \#}       & \textbf{$\kappa$}       \\ \hline
			1 & 0.67 & 6 & 0.91      \\ \hline
			2 & 0.74 & 7 & 0.87        \\ \hline
			3 & 0.82 & 8 & 0.90    \\ \hline
			4 & 0.84 & 9 & 0.94 \\ \hline
			5 & 0.84 & 10 & 0.91     \\ \hline
		\end{tabular}
		  \caption{Agreement in different stages}
		  \label{tab:kappa-calc}
	\end{subtable}%
	\caption{Labeling agreement calculation}
\end{figure}

Furthermore, after finishing labeling the pipelines stages, we also classified the subject references into four classes based on the overall purpose of the article. 
First, after a few discussions, the raters and moderator came up with the classes. 
Then, the raters classified each pipeline into one class. 
We found disagreements in 6 out of 71 references, which the moderator reconciled with separate meetings with the two raters.
Based on our labeling, the literature that we collected are divided into four classes: describe or propose DS pipeline, survey or review, DS optimization, and introduce new method or application. 
Next, we are going to discuss the result of analyzing the DS pipelines in theory.

\subsection{Representative Pipeline in Theory}
The labeled pipelines with their stages are visually illustrated in the artifact \tabref{tab:label}.
We found that pipelines in theory can be both software architecture and \textit{team processes} unlike pipelines in-the-small and in-the-large. Through the labeling process, we separated those team processes (25 out of 71), which are discussed in \secref{subsec:characteristics-in-theory}.

\begin{figure*}[h!t]
	\centering
	\includegraphics[width=\linewidth]{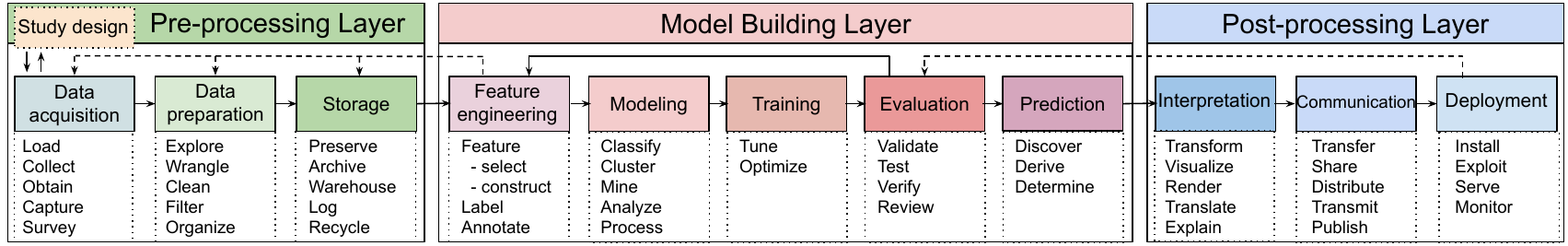}
	\caption{Concepts in a data science pipeline. The sub-tasks are listed below each stage. The stages are connected with feedback loops denoted with arrows. Solid arrows are always present in the lifecycle, while the dashed arrows are optional.
	Distant feedback loops (e.g., from \textit{deployment} to data \textit{acquisition}) are also possible through intermediate stage(s).
	}
	\label{lc}
\end{figure*}

\begin{table}[h!tb]
  \centering
  \footnotesize
    \begin{tabular}{|p{8.15cm}|}
    \hline
    \rowcolor[rgb]{ .251,  .251,  .251} \multicolumn{1}{|c|}{\textcolor[rgb]{ 1,  1,  1}{\textbf{Stages of Data Science Pipeline}}} \\
    \hline
    \textbf{Data Acquisition (ACQ):} In the beginning of DS pipeline, data are collected from	appropriate sources. Data can be acquired manually or automatically. Data acquisition also involves understanding the nature of the data, collecting relevant data, and integrating available datasets. \\
    \hline
    \rowcolor[rgb]{ .949,  .949,  .949} \textbf{Data Preparation (PRP):} Data are generally acquired in a raw format that needs certain preprocessing steps. This involves exploration and filtering, which helps identify the correct  data for further processing. Well prepared data reduces the time required for data analysis and contributes to the success of the DS pipeline. \\
    \hline
    \textbf{Storage (STR):} It is important to find an appropriate hardware-software combination to preserve data so that it can be processed efficiently. For example, \citeauthor{miao2017provdb} used graph database system Neo4j  \cite{miller2013graph} to build a collaborative analysis pipeline \cite{miao2017provdb}, since Neo4J supports querying graph data properties. \\
    \hline
    \rowcolor[rgb]{ .949,  .949,  .949} \textbf{Feature Engineering (FTR):} The entire dataset might not contribute equally to decision making. In this stage, appropriate features that are useful to build the model are identified or constructed. Features that are not readily available in the dataset, require engineering to create them from raw data.\\
    \hline
    \textbf{Modeling (MDL):} When data are preprocessed and features are extracted, a model is built to analyze the data. Model building includes model planning, model selection, mining and deriving important properties of data. Appropriate data processing strategies and algorithms are selected to create a good model. \\
    \hline
    \rowcolor[rgb]{ .949,  .949,  .949} \textbf{Training (TRN):} For a specific model, we need to train the model with available labeled data. By each training iteration, we optimize the model and try to make it better. The quality of the training dataset contributes to the training accuracy of the model. \\
    \hline
   \textbf{ Evaluation (EVL):} After training the model, it is tested with a new dataset which has not been used as training data. Also, the model can be evaluated in real-life scenarios and compared with other competing models. Existing metrics are used or new metrics are created to evaluate the model.\\
    \hline
    \rowcolor[rgb]{ .949,  .949,  .949} \textbf{Prediction (PRD):} The success of the model depends on how good a model can predict in an unknown setup. After a satisfactory evaluation, we employ the model to solve the problem and see how it works. There are many prediction metrics such as classification accuracy, log-loss, F1-score, to measure the success of the model.  \\
    \hline
    \textbf{Interpretation (INT):} The prediction result might not be enough to make a decision. We often need a transformation of the prediction result and post-processing to translate predictions into knowledge. For example, only numerical results do not help much but a good visualization can help to make a decision. \\
    \hline
    \rowcolor[rgb]{ .949,  .949,  .949} \textbf{Communication (CMN):} Different components of the DS system might reside in a distributed environment. So, we might need to communicate with the involved parties (e.g., devices, persons, systems) to share and accumulate information. Communication might take place in different geographical locations or the same. \\
    \hline
    \textbf{Deployment (DPL):} The built DS solution is installed in its problem domain to serve the application. Over time, the performance of the model is monitored so that the model can be improved to handle new situations. Deployment also includes model maintenance and sending feedback to the model building layer. \\
    \hline
    \end{tabular}%
  \vspace{5mm}
  \caption{Description of the stages in DS pipeline}
  \vspace{-6mm} 
  \label{tab:stages-description}
\end{table}

\textbf{RQ1a: What is a representative definition of the DS pipe\-line in theory?}
From the empirical study, we created a representative rendition of DS pipeline with 3 layers, 11 stages and possible connections between stages as shown in \fignref{lc}. Each shaded box represents a DS stage that performs certain sub-tasks (listed under the box). In the preprocessing layer, the stages are \textit{data acquisition}, \textit{preparation}, and \textit{storage}.
The preprocessing stage \textit{study design} only appeared in team process pipelines that comprise requirement formulation, specification, and planning, which are often challenging in data science.
The algorithmic steps and data processing are done in the model building layer.
\textit{Modeling} does not necessarily imply the existence of an ML component, since DS can involve custom data processing or statistical modeling.
Post-processing layer includes the tasks that take place after the results have been generated.
The DS pipeline stages are described in \tabref{tab:stages-description}. 

\begin{figure}[]
	\centering
	\includegraphics[width=0.95\columnwidth]{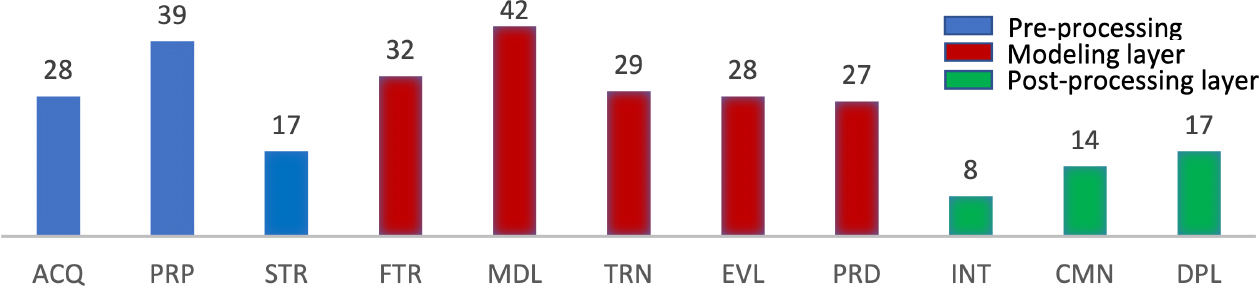}
	\caption{Frequency of pipeline stages in theory}
	\label{freq}
\end{figure}

\textbf{RQ1b: What are the frequent and rare stages of the DS pipeline in theory?}
The frequency of stage can depend on the focus of the pipeline or its importance in certain context (ML, big-data management).
Among 46 DS pipelines (which are not team processes), \fignref{freq} shows the number of times each stage appears. 
A few pipelines present stages with broad terminology that fit multiple stage-definitions. In those cases, the pipelines were labeled with the fitted stages and counted multiple times.
\textit{Modeling, data preparation, and feature engineering} appear most frequently in the literature. 
While \textit{modeling} is present in 93\% of the pipelines, other
model related stages (\textit{feature engineering, training, evaluation, prediction}) are not used consistently.
Often \textit{training} is not considered as a separate stage and 
included inside the \textit{modeling} stage. 
Similarly, we found that \textit{evaluation} and \textit{prediction} are 
often not depicted as separate stages.
However, by separating the stages and modularizing the tasks, the DS 
process can be maintained better~\cite{msml, sculley2015hidden}. 
The pipeline created with the most number of stages (11) is provided 
by~\citeauthor{ashmore2019assuring}~\cite{ashmore2019assuring}. 
On the other hand, about 15\% of the pipelines from the literature are 
created with a minimal number (3) of stages. Among them, 80\% are ML processes and falls in the category of DS optimizations. 
We found that these pipelines are very specific to particular applications, which include context-specific stages like data sampling, querying, visualization, etc., but do not cover most of the representative stages. 
A pipeline in theory may not require all representative stages, since it can have novelty in certain stages and exclude the others. However, the representative pipeline provides common terminology and facilitate comparative analysis.

\findings{Post-processing layers are included infrequently (52\%) compared to pre-processing (96\%) and model building (96\%) layers of pipelines in theory.}
Clearly, preprocessing and model building layers are considered in
almost all of the studies. In most of the cases, the pipelines do not consider
the post-processing activities (\textit{interpretation, communication, deployment}). 
These pipelines often end with the predictive process and thus do 
not follow up with the later stages which entails how the result 
is interpreted, communicated and deployed to the external environment. 
\citeauthor{miao2017ds} argued that overall lifecycle management tasks (e.g., model versioning, sharing) are largely ignored for deep learning systems \cite{miao2017ds}.
Previous studies also showed that significant amount of cost and effort is 
spent in the post-development phases in traditional software 
lifecycle~\cite{lientz1978characteristics, rajlich2014software}. 
In data-intensive software, the maintenance cost can go even higher 
with the high-interest technical debt in the pipeline~\cite{sculley2014machine}.
Therefore, post-processing stages should be incorporated for a better 
understanding of the impact of the proposed approach on maintenance of the DS pipeline.

\subsection{Organization of Pipeline Stages in Theory}

\textbf{RQ2: How are pipeline stages connected to each other?}
In \fignref{lc}, for simplicity, we depicted the DS pipeline as
a mostly linear chain. 
However, our subject DS pipelines often have non-linear behavior. 
In any stage, the system might have to return to the previous stage for 
refinement and upgrade, e.g., if a system faces a real-world challenge 
in \textit{modeling}, it has to update the algorithm which might affect 
the data pre-processing and feature engineering as well. 
Furthermore, the stages do not have strict boundaries in the DS lifecycle.  
In \fignref{lc}, two backward arrows,
from \textit{feature engineering} and \textit{evaluation}, indicate feedback to 
any of the previous stages. 
Although in traditional software engineering processes (e.g., waterfall model, 
agile development, etc.), feedback loop is not uncommon, in DS lifecycle, 
there are multiple stakeholders and models in a distributed environment which 
makes the feedback loops more frequent and complex. 
\citeauthor{sculley2015hidden} pointed that DS tasks such as sampling, learning, 
hyperparameter choice, etc. are entangled together so that Changing Anything 
Changes Everything (CACE principle)~\cite{sculley2015hidden}, which in turn 
creates implicit feedback loops that are not depicted in the pipelines~\cite{van2017versioning, chen2014data, ur2016big, gandomi2015beyond}.
The feedback loops inside any specific layer are more frequent than the 
feedback loops from one layer to another. 
Also, a feedback loop to a distant previous stage is expensive. 
For example, if we do \textit{data preparation} after \textit{evaluation} then 
the intermediate stages also require updates.

\subsection{Characteristics of the Pipelines in Theory}
\label{subsec:characteristics-in-theory}

\textbf{RQ3: What are the different types of pipelines available in theory?}
The context and requirements of the project can influence pipeline design and architecture \cite{garcia2018context}. Here, we present the types of pipelines with different characteristics that are available in theory.
We classified each subject in our study into four classes based on the overall goal of the article. 
The most of the pipelines in theory (39\%) are \textit{describing or proposing} new 
pipelines to solve a new or existing problem. 
About 31\% of the pipelines are on \textit{reviewing or comparing} the existing pipelines. 
The third group of DS pipelines (14\%) are intended to \textit{optimize} a certain 
part of the pipeline. 
For example,~\citeauthor{van2017versioning} proposes a pipeline for managing 
multiple versions of pipelines and optimize performance~\cite{van2017versioning}. 
Most of the pipelines in this category are application specific and include 
very few stages that are necessary for the optimization. 
Fourth, some research \textit{introduce new application} or method and present within the pipeline.
We observed that there is no standard methodology to develop comparable and 
inter-operable DS pipelines. 
Using the labeling methodology shown in \fignref{label}, 
we labeled each pipeline and found 
three types of DS pipelines in the literature: 1) ML process, 
2) big data management process, and 3) team process. 

\textit{ML process:} 46\% of all the pipelines we found in the literature are describing machine learning processes. The recent advent of artificial intelligence, supervised learning and deep learning has led to more DS systems that involve ML components. 
The pipelines in this category emphasize the algorithmic process, 
learning patterns, and building predictive models.
However, the post-processing stages are rare in these type of pipelines. 
The ML pipelines are often thought of as algorithmic process in 
the laboratory scenario. 
But as mentioned in~\cite{ashmore2019assuring}, incorporating the post-processing stages would be desired to ensure safe real-world deployment of such pipelines.

\textit{Big data management}: The references in this category
present DS pipelines that manage a large amount of data or describes a
framework (software-hardware infrastructure) for data processing but 
do not contain machine learning components in the pipeline. 
Processing large amount data often requires specific algorithms and 
engineering methods for efficiency and further processing. 
We found that 18\% of all the subject studies fall in this category. 

\textit{Team process:} We also found some DS pipelines that are not 
describing DS software architecture. 
These pipelines describe workflow of human activities that needs to be 
followed in a DS pipeline. 
These studies present a high-level view for building DS component in 
a team environment. 
The data science teams require specific expertise and management to 
build successful DS pipelines~\cite{kim2016emerging, msml}.
In this paper, in \secnref{sec:in-the-small} and \secnref{sec:in-the-large}, 
we are only focusing on DS pipeline as software architecture, and therefore, 
we did not compare the team process pipelines in the rest of this section.

\findings{Most of the pipelines in-theory involve cyber and physical components, only a few with human processes in the loop.}
We identified whether the pipelines involve cyber, physical or human process, using our labeling process described in 
section \secref{ldsp}. \textbf{Cyber} processes refer to activities that involve
automated systems and machinery computations. 
Since modern DS systems involves large amount of data and requires extensive 
computation, all of the pipelines include cyber component in it. 
\textbf{Physical} processes include the activities which require real-world 
connections with the system. 
For example, collecting data using mobile sensors or cameras is a physical process.
Although 23\% of the big data pipelines include physical processes, only 
9\% of the ML pipelines include that in the pipeline.
In many DS systems, developers or researchers participate in the pipelines
actively to make decisions that need \textbf{human} interventions~\cite{todd2017computing, van2017versioning}. 
For example, in many DS systems, analytical model validation, troubleshooting, 
data interpretation is necessary which requires human involvement. 
However, only 13\% of the pipelines acknowledged human involvement in the pipeline.

\section{DS Pipeline in-the-Small}
\label{sec:in-the-small}

Similar to the DS pipelines in large systems and frameworks, for a very
specific data science task (e.g., object recognition, weather forecasting, etc.),
programmers build pipeline. Different stages of the program perform a
specific sub-task and connect with the other stages using data-flow or control-flow relations.
In this section, we described such DS pipelines
\textit{in-the-small}. 

\subsection{Methodology}

We collected 105 DS programs from Kaggle competition notebooks \cite{kaggle}. 
Kaggle is one of the most popular crowd-sourced platforms for DS competitions, owned by Google. Besides participating in competitions, data scientists, researchers,
developers collaborate to learn and share DS knowledge in variety of domains.
The users and organizations can host a DS competition in Kaggle to solve real-world problems. 
A competition is accompanied by a dataset and prize money. 
Many Kaggle solutions have resulted in impactful DS algorithms and research such as neural networks used by Hinton and Dahl \cite{dahl2014multi}, improving the search for the Higgs Boson at CERN \cite{higs-boson}, etc. 
We chose Kaggle solutions to analyze DS pipeline for three reasons: 
1) all programs perform a DS task and provide solution to a well specified problem associated with a dataset,
2) solutions with the highest number of votes are well accepted solutions for a specific problem, and 
3) the problems cover a wide range of domains.

\begin{figure}[t]
	\centering
	\includegraphics[width=\columnwidth]{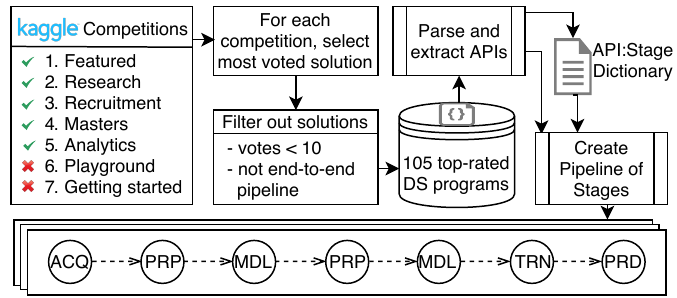}
	\caption{The pipeline creation process for Kaggle programs}
	\label{pipe}
\end{figure}

There are 331 completed competitions in Kaggle to date. They categorized the competitions into \textit{Featured},
\textit{Research}, \textit{Recruitment}, \textit{Masters}, \textit{Analytics}, \textit{Playground} and \textit{Getting started}. We collected solutions of all the competitions from each category except
\textit{Getting started} and \textit{Playground} (these two categories are intended to serve as DS tutorials and toy projects). 
First, we filtered the competitions for which there are solutions available (many old competitions do not contain any public solution). We found 138 such competitions. For a given competition problem, we selected the most voted solution which has at least 10 votes. 
Thus, we got 105 top-rated DS solutions for analyzing pipelines in-the-small. This selection and pipeline creation process is shown in \fignref{pipe}.

All of the DS programs are written in Python using ML libraries like Keras, Scikit-learn, Tensorflow, etc. These packages provide high-level Application Programming Interfaces (APIs) for performing a specific task on data or model. We parsed the programs into Abstract Syntax Tree (AST) and collected all the API calls from the programs. 
Then the functionality of an API is used to identify the stage of the pipeline. 
We extracted the temporal order of API calls to identify the stages. Standard static analysis of the Python programs facilitate the extraction process. Our analysis suggests that the DS programs follow a linear structure with less than 4\% AST nodes being conditional or loops. \citeauthor{wang2021restoring} proposed a similar approach for extracting external dependencies in Jupyter Notebooks by creating an API database and analyzing AST \cite{wang2021restoring}.

We created a dictionary by mapping each API collected from the programs, to one of the 11 stages of the DS pipeline described in section \secnref{sec:theory}. During the mapping, we excluded the generic APIs from the dictionary. For example, \texttt{model.summary()} is used to print the model parameters and does not represent any stage of the pipeline. 
For creating the dictionary, we taken a two-fold approach. First, we
understand the context of the program and API usage. Second, we look at the API documentation to confirm the corresponding pipeline stage.
We found that DS APIs are definitive in their operations and well-categorized by the library. For example, the APIs in Keras \cite{keras-doc} and Scikit-learn \cite{sklearn-doc} are grouped into preprocessing, models, etc. Our API-dictionary was manually validated by a second-rater and moderator who labeled DS pipelines in section \secref{sec:theory}.
Then, we built a tool which takes the API dictionary and DS program, and automatically creates the DS pipeline. For a sequence of APIs with the same stage, we abstracted them into a single stage. As an example, \fignref{pipe} shows a DS
pipeline created from a Kaggle solution \cite{lstmx}. Each stage in the pipeline
(e.g., ACQ, PRP) represents one or more API usages. The arrows in the pipeline denote the temporal sequence of stages.
Note that, one stage can appear multiple times in a pipeline. 
The API dictionary, Kaggle programs, and tool to generate the pipelines is 
shared in the paper artifact \cite{sumon2021pipeline}.

\subsection{Representative Pipeline in-the-Small}

\begin{figure}[t]
	\centering
	\includegraphics[width=\columnwidth]{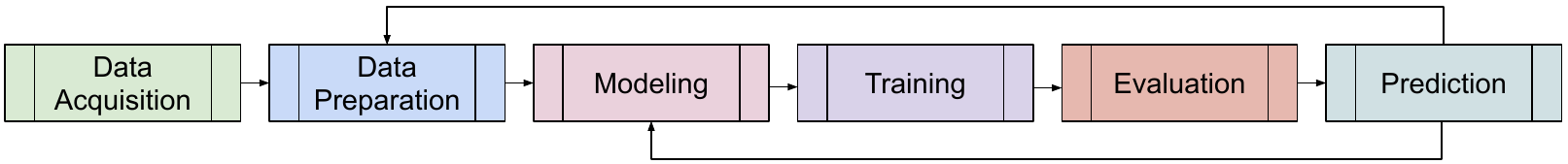}
	\caption{Pipeline \textit{in-the-small} extracted from API usages}
	\label{pipeline}
\end{figure}

\textbf{RQ4: What are the stages of \textit{DS pipeline in-the-small}?} 
Among the 11 pipeline stages described in \fignref{lc}, we
found only 6 stages in the DS programs that are depicted in
\fignref{pipeline}. 
Other stages (e.g., \textit{storage, feature engineering, interpretation, 
communication, deployment}) are not found in these programs because 
these stages occur while building a production-scale large DS system 
and often not present in the DS notebooks. Therefore, the pipeline in DS programs consists of the subset of pipeline stages in theory. 

\begin{figure}[t]
	\centering
	\includegraphics[width=.7\columnwidth]{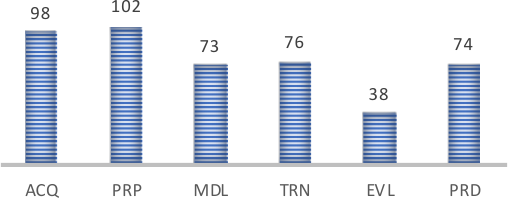}
	\caption{Frequency of pipeline stages \textit{in-the-small}}
	\label{freq2}
\end{figure}

We summarized the frequency of each stage of the DS programs in \fignref{freq2}. 
Among 105 programs, \textit{data acquisition} and \textit{data preparation} 
are present in almost all of them.
Surprisingly, \textit{modeling} is present in only 70\% of the programs. 
We found that, in many programs, no modeling APIs had been used 
because developers did not use any built-in ML algorithm from libraries, 
e.g., LogisticRegression, LSTM, etc. 
In these cases, the developers use data-processing APIs on the training 
data to build custom model, e.g., this notebook~\cite{rubase} uses 
\textit{data preparation} APIs to produce results. 
To enable more abstraction of the stages in these pipelines, further 
modularization is necessary, which has been investigated in RQ8.

\findings{\textit{Evaluation} stage is infrequent, appearing only in 36\% of the pipelines in-the-small.}
\textit{Evaluation} is a tricky stage of the DS pipeline. 
Developers have to choose the appropriate metric and 
methodology to evaluate their model. 
Based on the evaluation result, the model is updated over multiple iterations. 
We found that, besides using metrics, in many cases, evaluation 
requires human understanding and comparison of the result produced by the model.
The reason for having less number of \textit{evaluation} stage in the 
pipeline is that often the developers evaluate the performance by 
plotting and visualizing the result. 
Since the visualization APIs are not considered as \textit{evaluation} 
stage, we found this stage less frequently in pipelines. 
Also, many programs directly go to the \textit{prediction} stage without 
going to \textit{evaluation} stage at all. 
Furthermore, notebooks are often used for experimentation purposes so 
that many computations are performed during development but eliminated 
when the notebooks are shared~\cite{kery2018story}. 
For example, one developer might try a number of classifiers and evaluate 
their accuracy. 
After finding the best performing classifier, it can be the only one 
shared in the notebook.
Therefore, we experienced many missing stages in the pipeline in-the-small.
The complex DS tasks require several computations which might not be 
used in producing the final prediction, but definitely should be considered 
as part of the pipeline.

\subsection{Pipeline Organization in the Small}
\begin{figure}[]
	\centering
	\includegraphics[width=\columnwidth]{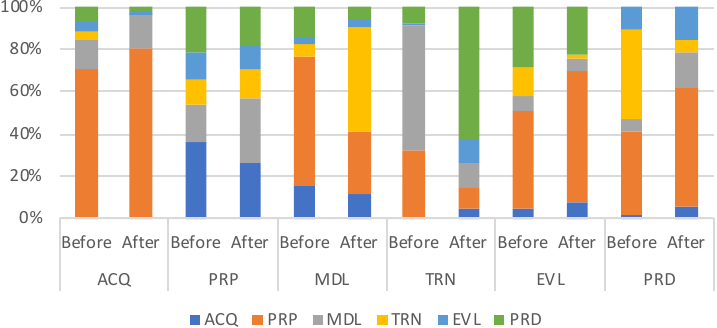}
	\caption{Stages occurring before and after each stage}
	\label{conn}
\end{figure}

\textbf{RQ5: How are the stages connected with each other in pipe\-line in-the-small?}
 To answer RQ5, we considered each occurrence of the stages in a DS
program and looked at its previous and next stage. In \fignref{conn}, we showed which stages are followed or preceded by each stage. We found
that \textit{data preparation} can occur before or after all other
stages. Apart from that, \textit{data acquisition} is followed by
\textit{data preparation} most of the time, which in turn is followed by \textit{modeling}. \textit{Modeling} is followed mostly by \textit{training}, which in turn is followed by \textit{prediction}. \textit{Evaluation} is mostly surrounded by
\textit{prediction} and \textit{data preparation}. From \fignref{conn}, we can also find some most occurring feedback loop: \textit{evaluation} to \textit{preparation}, \textit{evaluation} to \textit{modeling} and \textit{prediction} to \textit{modeling}.

\textit{Data preparation} tasks (e.g., formatting, reshaping, sorting) are not limited to just before the \textit{modeling} stage, rather it
is done on a \textit{whenever-needed} basis. 
For example, in the following code snippet from a Kaggle competition \cite{siamese}, while creating model-layers, data preprocessing API has been called in line 2. 

\begin{lstlisting}[language=Python]
x = Conv2D(mid, (4, 1), activation='relu', padding='valid')(x)
x = Reshape((branch_model.output_shape[1], mid, 1))(x)
x = Conv2D(1, (1, mid), activation='linear', padding='valid')(x)
x = Flatten(name='flatten')(x)
head_model = Model([xa_inp, xb_inp], x, name='head'
\end{lstlisting}

The \textit{modeling} stage is always surrounded by other stages of the pipeline.
However, there is often a loop around \textit{modeling, training, evaluation, and prediction}. \textit{Modeling} often repeats many times to improve the model over multiple iterations.
For example, in the following Kaggle code snippet \cite{willko}, the model is created and trained multiple times to find the best one.
\begin{lstlisting}[language=Python]
random_forest = RandomForestClassifier(n_estimators=100, random_state=50, verbose=1, n_jobs=-1) # Modeling
random_forest.fit(train, train_labels) # Train
...
poly_features = scaler.fit_transform(poly_features) 
poly_features_test = scaler.transform(poly_features_test)
random_forest_poly = RandomForestClassifier(n_estimators=100, random_state=50, verbose=1, n_jobs=-1) # Modeling
random_forest_poly.fit(poly_features, train_labels) # Training
pred = random_forest_poly.predict_proba(poly_features_test)[:,1]
\end{lstlisting}
\begin{figure*}[t]
	\centering
	\includegraphics[width=\linewidth]{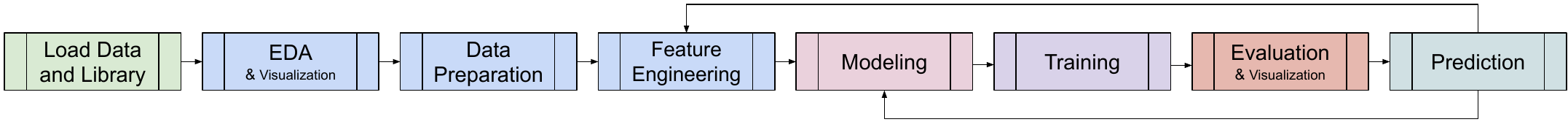}
	\caption{Representative data science pipeline \textit{in-the-small}}
	\label{pipeline-updated}
\end{figure*}
\findings{Stages of pipelines in-the-small are often tangled with each other.}
All of the DS programs fail to maintain a good separation of 
concerns~\cite{dijkstra1982role} between stages. 
Strong abstraction boundaries help to make the program modular and 
easy-to-maintain~\cite{parnas1985modular, pan20decomposing, pan22decomposing}. 
In addition, a good DS solution should not only compute better 
predictive result, but also facilitate software engineering activities e.g., debugging, testing, monitoring~\cite{hill2016trials}. 
However, we found that stages are often {\em tangled} with other stages~\cite{Kiczales97,Prehofer97,CalderKolbergMagill03} across the pipelines.
The code for one stage is interspersed with the code for other stages.
For example, while building the deep learning network (\textit{modeling}), 
the developers often switch to different \textit{data preparation} tasks, 
e.g., reshaping, resizing \cite{islam19, islam20repairing}, which tangles data preparation 
concern with the modeling concern.
We observed some early attempts to adopt modular design practices.
For instance, this notebook \cite{danielbecker} separated code into 
different high-level stages, namely, \textit{preparation}, \textit{feature extraction}, 
\textit{exploratory data analysis (EDA)}, \textit{topic model}, etc. 
These high-level pipelines can improve the abstraction, which 
further enable the maintainability, and reusability~\cite{rule2018exploration}. 
In some scenarios, reuse or maintenance might not be desired for pipelines in-the-small. However, to enhance readability \cite{kery2018story} and repeatability \cite{hill2016trials} and ease of testing, debugging or repairing \cite{wardat21deeplocalize, wardat2021deepdiagnosis},
more attention on modular design practices is needed for DS pipelines.

\findings{Data preparation stage is occurring significant number of times between any two stages of pipelines in-the-small, which is causing pipeline jungles.}
We found that new data sources are added, new features are identified, and new values are calculated incrementally in the pipeline which evolves organically. 
This results in a large number of data preprocessing tasks like sampling, joining, resizing along with random file input-output. This is called \textit{pipeline jungles}~\cite{sculley2015hidden}, which causes technical debt for DS systems in the long run. 
\textit{Pipeline jungles} are hard to test and any small change in the pipeline will take a lot of effort to integrate. 
The situation gets worse in case of larger DS pipelines, where several data management activities (e.g., clean, serve, validate) are necessary through the pipeline in different stages \cite{polyzotis2017data, polyzotis2018data}.
The recommended way is to think about the pipeline holistically and scrape the pipeline jungle by redesigning it, which in turn takes further engineering effort~\cite{sculley2015hidden}. 
We found that the large DS projects, which are discussed in 
\secref{sec:in-the-large}, isolate the data preparation tasks into separate files and modules~\cite{cnnt, darkf, facenet, qanett}, which alleviates the pipeline jungles problem. 
So, DS pipeline in-the–small needs further IDE (e.g., Jupyter Notebook, etc.) support and methodologies for code isolation and modularization.

\subsection{Characteristics of Pipelines in-the-Small}
\textbf{RQ6: What are the patterns in pipeline in-the-small and how it compares to pipeline in theory?}
We have not found many stages from \fignref{lc}, e.g., feature engineering, 
interpretation, communication, in pipeline in-the-small. 
One reason is that the low-level pipeline extracted from the API usages cannot 
capture some stages. 
For example, even if a developer is conducting feature engineering, 
the used APIs might be from the data preparation stage. 
Fortunately, we found many Kaggle notebooks that are organized by the pipeline stages.
We visited all the 105 Kaggle notebooks in our collection and 
extracted these high-level pipelines manually. 
Unlike the low-level pipelines (extracted using API usages), a 
high-level pipeline consists of the stages abstracted by the developers.

The Kaggle notebooks follow literate programming paradigm~\cite{wagner2020accountability, rule2018exploration}, 
which allows the developers to describe code using rich text and 
separate them into sections. 
We found that 34 out of 105 notebooks divided the code into stages. 
We collected those stages from the Kaggle notebooks. 
Furthermore, we labeled these notebooks into the 11 stages from 
DS pipeline in theory by two raters, and extracted the stages that 
are not present in theory. 
The extracted high-level pipelines and labels are available in the 
paper artifact~\cite{sumon2021pipeline}.

We observed that no notebooks specify these stages: \textit{storage, 
interpretation, communication,} and \textit{deployment}. 
These DS programs are not production-scale projects. 
Therefore, they do not include the post-processing stages in the pipeline. 
The most common stages are \textit{modeling} (79\%), 
\textit{data preparation} (62\%), \textit{data acquisition} (53\%), 
and \textit{feature engineering}(35\%), which is aligned with the 
finding of DS pipeline in theory. 
In addition, we found these stages which are not present 
in theory: \textit{library loading, exploratory data analysis (EDA), 
visualization}.
Among them \textit{EDA} has been used most of the times (43\%) and covered the most part of those pipeline. 
Before going to the modeling and successive stages, a lot of effort 
is given on understanding the data, compute feature importance, and 
visualize the patterns, which help to build 
models quickly in later stages \cite{ashmore2019assuring}.

Furthermore, some notebooks present \textit{library loading} as separate stage. 
We observed that choosing appropriate library/framework and setting 
up the environment is an important step while developing pipeline in-the-small.
We also found that data visualization is an recurring stage mentioned 
by the developers. 
Visualization can be done for EDA or feature engineering (before modeling), or for evaluation (after modeling). 
Based on these observations we updated the representative pipeline 
in-the-small in \fignref{pipeline-updated}.
The high-level pipeline provides an overall representation of the system, which can be leveraged to design software process. It would be beneficial for the developers to close the gap between the low-level and the high-level pipeline by identifying the tangled stages.
\section{DS Pipeline in-the-Large}
\label{sec:in-the-large}

The DS solutions described in the previous section are specific to a given dataset and a well-defined problem.
However, there are many DS projects which are large, not limited to a single source file, and contains multiple modules.
These solutions are intended to solve more general problems which might not be specific to a dataset. 
For example, the objective of the \textit{Face Classification} project in \gh \cite{face} is to detect face from images or videos and classify them based on gender and emotion. 
This problem is not specific to a particular dataset and the scope is
broader compared to the Kaggle solutions. We collected such top-rated DS projects from \gh to analyze DS pipeline in-the-large.

\subsection{Methodology}
\citeauthor{biswas2019boa} published a dataset containing
top rated DS projects from \gh \cite{biswas2019boa}.
From the list of projects in this dataset, we filtered mature DS projects
having more than 1000 stars. Thus, we found 269 mature \gh projects.
However, there are many projects in this list which are DS libraries,
frameworks or utilities. Since we want to analyze the pipeline of data science software, we removed those projects.
Finally, we also removed the repositories which serve educational purposes.
Thus, we found a list of 21 mature open-source DS projects.
The list of projects, and their purpose are shown in \tabref{tab:gh-table}.

\begin{table}[t]
	\vspace{12pt}
	\scriptsize
	\setlength\tabcolsep{1.2pt}
	\centering
	\caption{GitHub projects for analyzing pipeline in-the-large}
	  \begin{tabular}{|p{2.48cm} p{3.81cm} r r r|}
	  \hline
	  \rowcolor[rgb]{ .251,  .251,  .251} \textcolor[rgb]{ .949,  .949,  .949}{\textbf{Project Name}} & \textcolor[rgb]{ .949,  .949,  .949}{\textbf{Purpose}} & \multicolumn{1}{l|}{\textcolor[rgb]{ .949,  .949,  .949}{\textbf{\#Files}}} & \multicolumn{1}{l|}{\textcolor[rgb]{ .949,  .949,  .949}{\textbf{\#AST}}} & \multicolumn{1}{l|}{\textcolor[rgb]{ .949,  .949,  .949}{\textbf{LOC}}} \\
	  Autopilot \cite{autopilot} & Pilot a car using computer vision & 36    & 11185 & 348 \\
	  \rowcolor[rgb]{ .949,  .949,  .949} CNN-Text-Classification \cite{cnnt} & Sentence classification & 69    & 47797 & 11.4K \\
	  Darkflow \cite{darkf} & Real-time object detection and classification & 1025  & 655670 & 8.6K \\
	  \rowcolor[rgb]{ .949,  .949,  .949} Deep ANPR \cite{anpr} & Automatic number plate recognition & 64    & 70464 & 10.8K \\
	  Deep Text Corrector \cite{tcorr} & Correct input errors in short text & 47    & 50770 & 3.0K \\
	  \rowcolor[rgb]{ .949,  .949,  .949} Face Classification \cite{face} & Real-time face and emotion/gender detection & 292   & 117901 & 35.3K \\
	  FaceNet \cite{facenet} & Face recognition & 1352  & 1889529 & 18.2K \\
	  \rowcolor[rgb]{ .949,  .949,  .949} KittiSeg \cite{kitti} & Road segmentation & 276   & 187143 & 4.8K \\
	  LSTM-Neural-Network \cite{lstm} & Predict time series steps and sequences & 24    & 11434 & 1.2K \\
	  \rowcolor[rgb]{ .949,  .949,  .949} Mask R-CNN \cite{matterport} & Object detection and instance segmentation & 256   & 1567786 & 15.6K \\
	  MobileNet SSD \cite{mobnet} & Object detection network & 28    & 21272 & 25.6K \\
	  \rowcolor[rgb]{ .949,  .949,  .949} MTCNN \cite{mtcnn} & Joint face detection and alignment & 153   & 121138 & 219.7K \\
	  Object-Detector-App \cite{odetect} & Real-time object recognition & 215   & 318534 & 47.9K \\
	  \rowcolor[rgb]{ .949,  .949,  .949} Password-Analysis \cite{passd} & Analyze a large corpus of text passwords & 148   & 67870 & 3.6K \\
	  Person Blocker \cite{personb} & Block people in images & 12    & 44517 & 977 \\
	  \rowcolor[rgb]{ .949,  .949,  .949} QANet \cite{qanett} & Machine reading comprehension & 83    & 107669 & 2K \\
	  Speech-to-Text-WaveNet & Sentence level english speech recognition & 32    & 18626 & 5.1K \\
	  \rowcolor[rgb]{ .949,  .949,  .949} Tacotron \cite{taco} & Text-to-speech synthesis & 114   & 58845 & 1.4K \\
	  Text-Detection-CTPN \cite{ctpn} & Text detection & 640   & 257083 & 18.4K \\
	  \rowcolor[rgb]{ .949,  .949,  .949} TF-Recomm \cite{recom} & Recommendation systems & 17    & 7789  & 535 \\
	  XLNet \cite{xlnet} & Language understanding & 36    & 143172 & 11.5K \\
	  \hline
	  \end{tabular}%
	  \label{tab:gh-table}
  \end{table}%

\begin{figure}[t]
	\centering
	\includegraphics[width=\columnwidth]{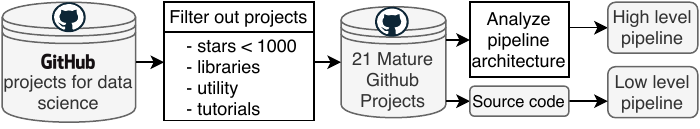}
	\caption{Pipeline creation process for \gh projects}
	\label{gh}
\end{figure}

For each project, we created two pipelines: high-level pipeline and low-level pipeline. For creating the high-level pipeline, we manually checked the project architecture, module structure and execution process. This gave us a good understanding of the source file organization and linkage between modules. 
After identifying the high-level pipeline and execution sequences of the source files, we used the same API based method used to analyze Kaggle programs in the previous section, to create low-level pipeline of these \gh projects. The methodology of selecting and extracting pipelines from the GitHub projects is shown in \fignref{gh}.

For example, the project \textit{QANet}~\cite{qanett} is intended to do machine reading
comprehension. Here, Python has been used as the primary language, and
shell script has been used for data downloading and project setup. 
The high-level pipeline for \textit{QANet} includes the stages: 
\textit{data acquisition}, \textit{data preparation}, \textit{modeling}, 
\textit{training}, \textit{evaluation} and \textit{prediction}. 
In the beginning, \texttt{config.py} file integrates the modules (preparation, modeling, and training) and provides an interface to
configure a model by specifying dataset and other parameters. 
Then, the file \texttt{evaluate.py} is executed to perform the evaluation 
and prediction. 
For the low-level pipeline, for a specific file, we used the API based 
analysis to generate the pipeline, which was used to analyze pipeline 
in-the-small. 
For instance, in the project \textit{QANet}, although
\texttt{model.py} serves modeling at a high level, it also does data
preparation, training, and evaluation, when APIs are considered.
In addition to the pipeline stages, we also identified a few other properties of each project:
1) number of contributors,
2) AST count,
2) technology/language used,
3) entry points and
4) execution sequence.
We leveraged the Boa infrastructure~\cite{Dyer-Nguyen-Rajan-Nguyen-13, dyer2015boa-b} 
to analyze the different properties of the projects.
These properties helped us to categorize and analyze the pipeline in-the-large. 
The details of the projects are available in the paper artifact~\cite{sumon2021pipeline}.

The projects are from various domains: object detection, face classification,
automated driving, speech synthesis, number plate recognition, predict time
series sequence, etc. The number of developers in each project ranges 
between 1 and 40 with an average of 8. 
Among 21 projects, 16 of them are developed by teams and 5 of them are 
developed by individuals. 
The primary language used to develop these projects is Python.

\subsection{Representative Pipeline in-the-Large}
Compared to the Kaggle programs, we found a significant difference in the 
pipeline of large DS projects. 
Because of the larger size of the projects, the pipeline architecture is different. 
All the projects contain multiple source files for handling different tasks 
(e.g., modeling, training) and about 50\% of the projects organize the source 
files into modules (e.g., \textit{utils}, \textit{preprocessing}, \textit{model}, etc.).

\begin{figure}[t]
	\centering
	\includegraphics[width=.9\columnwidth]{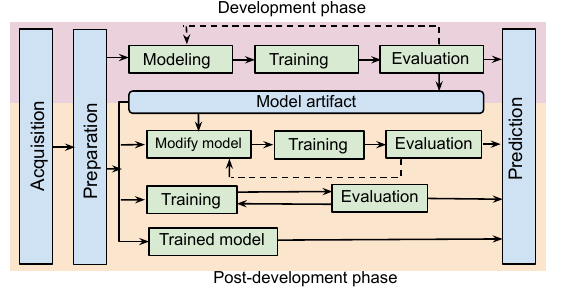}
	\caption{DS pipeline in-the-large. Development phase (top) runs during model building and post-development phase (bottom) runs for making prediction.
	}
	\label{gh-pipe}
\end{figure}

\textbf{RQ7: What is the representative DS pipeline in-the-large?}
Each of the projects contains six stages described in
\fignref{pipeline}: \textit{acquisition, preparation, modeling, training, evaluation}, and
\textit{prediction}. However, since the projects are not coupled to a specific dataset and they solve a more general problem, the projects are not limited to one single pipeline. We found that the pipeline of each project is divided into two phases: 1) development phase and
2) post-development phase, which is depicted in \fignref{gh-pipe}. 

In \textbf{development phase}, the main goal is to build a model that solves the problem in general. A base dataset is used to build the model that would be used for other future datasets. After completing a \textit{modeling, training, evaluation} loop, the final model is created and saved as an artifact. Afterwards, the projects also create model interfaces, which lets the user modify and exploit the model in the post-development phase. Finally, the model artifact is saved as a source file or some model archiving formats. For example, the project \textit{Person-Blocker} \cite{personb} and
\textit{Speech-to-Text-WaveNet} \cite{wavenet} saved the model in the source file (\texttt{model.py}) and lets the users train the model in the next phase.
On the other hand, the project \textit{KittiSeg} \cite{kitti} and \textit{Autopilot} \cite{autopilot} saved the built model artifact in JSON format (\texttt{.json}) and checkpoint format (\texttt{.ckpt}) respectively.
We observed that the evaluation and prediction is often not the main goal in this phase; rather, building an appropriate model and making it available for further usage is the central activity.

In \textbf{post-development phase}, the users access the pre-built model and use that for prediction. After acquiring data, a few preprocessing steps are needed to feed the model. In all of the projects under this study, we found that the development phase is similar. However, we identified three different patterns in the post-development phase which are shown in \fignref{gh-pipe}.
First, the users can modify the model by setting its hyperparameters and use that to make prediction on a new dataset.
Second, the users can use the model as-it-is and train the model on the new dataset to make prediction.
Third, the users can also download the pre-trained model and directly leverage that for prediction.
Finally, at the end of this phase, the prediction result is obtained.

The post-development phase in the pipeline enabled software reusability of the models. All of these projects have instructions in their \texttt{readme} or documentation explaining the usage and customization. For example, the project Deep ANPR \cite{anpr} provides instructions for obtaining large training data, retraining the models, and build it for prediction. However, not all the projects enable reusability in the development pipelines. Only a few of them provides access to the modules by importing in new development scenario. For instance, Darkflow \cite{darkf} let users access the \texttt{darkflow.net.build} module and use it in new application development. To increase the reusability of DS programs, it would be desired to consider similar access to the development pipeline of these large projects.

\subsection{Organization of DS Pipeline in-the-Large}
\textbf{RQ8: How are the stages connected in pipeline in-the-large?}
The abstraction in DS projects is stricter than the DS programs described in \secnref{sec:in-the-small}.
The projects are built in a modular fashion, i.e., one source file for a broad task (e.g., \texttt{train.py, model.py}). However, inside one specific file, there are many other possible stages, especially data \textit{preprocessing} appears inside all the source files.
In addition, the module connectivity is not linear. All of the modules use external libraries for performing different tasks. As a result, there are a lot of interdependencies (both internal and external) in the DS pipeline.
One immediate difference of these pipelines with traditional software is DS pipelines are heavily dependant on the data. For example, the project \textit{Speech-to-Text-WaveNet} \cite{wavenet} requires a certain format of data. When we want to use that in a new situation, the data properties might be different. So, the usage pipelines would have a few additional stages. In some cases, the original pipeline is modified. Here, there are many sub-pipelines work together to build a large pipeline. However, we have not found any framework or common methodology these software are using. The different patterns of DS pipelines seek more advanced methodology or framework to build DS pipeline and release for production.

\subsection{Characteristics of Pipelines in-the-Large }
\textbf{RQ9: What are the patterns found in the pipelines?}
The pipe\-lines found in this setting can be categorized into 1) loosely coupled and 2) tightly coupled, based on their modularity. 
A high number of contributors in the project resulted in loosely coupled pipelines. We found the loosely coupled ones are designed in a modular fashion and one module (e.g., data cleaning, modeling) is designed to be used by other modules.
Usually, there are multiple entry-points in a loosely coupled pipeline and user has more flexibility. On the other hand, in a tightly coupled pipeline, the modules are stricter and integrated tightly with other modules. There is only one or two entry-points to the pipeline, which automatically calls the other modules. We found that the projects with 6 or more contributors ($\sim$75\%) followed a loosely coupled architecture and projects with 1 to 5 contributors followed a tightly coupled architecture.

\findings{There is need for integration and deployment tools for pipelines in-the-large but no common framework is used in practice.}

Although all the project under this study are written using Python, no project is using any common tool that integrates the DS modules and provides interface to the pipeline.
Today, continuous integration and deployment (CI/CD) tools are widely used in traditional software lifecycle to automate compilation, building, and testing \cite{hilton2017trade, karlavs2020building}.
Additionally, from our subject studies of pipelines in theory, we found some CI/CD tools designed for ML pipelines available \cite{valohai2020pipeline, microsoft, databrick}. Surprisingly, here we found no projects in pipeline in-the-large are using any CI/CD tools. 
However, the projects demonstrate the need of CI/CD in the repositories.
In most of the projects, the environment setup and access to functionalities are configured through command lines scripts \cite{autopilot, ctpn}. Some projects used \textit{docker container} \cite{facenet, face, personb, wavenet} to set up the environment and run the pipeline. A few others used Python notebooks that call different modules to integrate the pipeline stages \cite{odetect, matterport, tcorr}. 7 out of 21 projects used shell script for integration (e.g., sending HTTP request to download data, model reuse, etc.) \cite{qanett, mobnet}.
Although CI/CD frameworks e.g., TravisCI,  GitHub Actions, Microsoft Azure DevOps are well established for traditional software such as web applications, several challenges remain for DS pipelines.
\citeauthor{karlavs2020building} outlined the probabilistic nature of ML testing as a major CI/CD challenge and pointed out the gap between recent theoretical development of CI/CD in DS and their usage in practice \cite{karlavs2020building}. Hence, further research is needed to investigate the practical challenges of using CI/CD in data science projects.
\section{Discussion}
\label{sec:discussion}
Through our survey, empirical study, and analysis, we presented the state of data science pipeline that describes its semantics, design concerns, and the overall computational paradigm. Furthermore, the findings show the importance of studying the pipeline structure reminiscing the traditional software engineering works on design patterns and architecture. 

\textit{In Theory:}
We presented all the representative stages and subtasks that inform the terminology of DS pipelines to be used in future works. By comparing with the available pipeline categories e.g., ML process, big data, and team processes, similarities and divergences can be directly identified. The presence of implicit feedback loops and lack of post-processing stages suggest ad hoc pipeline construction at the present time. This paper takes the first step towards comparable and reusable pipeline construction. 

\textit{In-The-Small:}
The novel API-based analysis can be utilized for mining, extracting, and statically analyzing pipelines.
We also elicited the notion of high-level and low-level pipelines, where the high-level abstraction has more similarity with that in theory. However, low-level pipelines exhibit many differences such as missing some stages, sparse data preparation, lack of modularization. The gap between low-level pipeline and its presentation in high-level can be reduced by making pipeline specific features available in development environment e.g., pipeline template in Jupyter Notebook. Additionally, the low-level pipelines often had an important stage \textit{exploratory data analysis} missing which incurs much time and effort. Pipeline versioning techniques that consider data, model, and source code will facilitate storing such intermediate stages.

\textit{In-The-Large:}
Different pipeline patterns emerged in development and post-development phase of the large projects, which suggest creating separate \textit{developer-centric} and \textit{user-centric} pipeline structure. 
In tightly-coupled projects, the abstraction of stages are contingent upon the project-specific requirements and internal/external dependencies, whereas, in loosely-coupled projects, opportunities remain to build reusable sub-pipelines that span over project boundaries. Finally, there is a need for building automated CI/CD tools for data science specific testing, deployment, and maintenance.

\textbf{To researchers and tool builders.}
(1) Modularization of DS pipeline into stages is challenging over all three representations. Further works are needed for standardization of pipeline architecture e.g., defining the interfaces of stages, enumerating externally visible properties, identifying domain-specific constraints, to develop reusable and interoperable DS pipelines.
(2) We showed potentials for automatic pipeline analysis framework based on static analysis and API specifications. A few future directions would be mining (sub-)pipelines patterns, build AutoML pipelines \cite{nguyen2021manas}, and analyzing evolution.
(3) We confirmed several antipatterns of pipelines that call for actions e.g., CACE principle, pipeline jungles, scarce post-processing, implicit feedback loop, CI/CD challenges.
(4) Pipeline specific tool support is needed such as version control for data and models, storing intermediate results between stages.

\textbf{To data scientists and engineers.}
(1) Pipelines are often built for a prototype in-the-small, which might not scale to a production level system. A well-designed pipeline in the early stage will help to identify key components, estimate cost, optimize, and manage risks better in the lifecycle.
(2) The representative views of pipelines will serve as a checklist of stages and their connections. 
(3) Data and algorithms being the focus of DS pipeline, preprocessing and modeling activities are well understood and practiced by data scientists. However, they should emphasize more on including rigorous \textit{evaluation} beyond accuracy such as robustness and fairness \cite{biswas20machine,biswas21fair}.
(4) Many people with diverse backgrounds are involved in a DS pipeline. A pipeline with human-in-the-loop approach will benefit identifying collaboration points, decomposing tasks, and manage transdisciplinary teams. For example, a pipeline can encourage data scientists to choose a modeling technique that is maintainable.
(5) Future work is necessary to identify the interactions of DS pipeline with the real world i.e., which stages receive inputs, when a checkpoint is saved, how results are disseminated, etc.
\section{Threat to Validity}
\label{sec:threat}
For building the pipelines from DS programs, we relied on the APIs. One
threat might be, what happens if the developer does not use any API for
completing a stage in the program. We examined this possibility and found
that DS programs are heavily dependent on libraries and external APIs and ML
tasks are always performed using library APIs. 
Additionally, we validated the API-to-stage dictionary with the API documentation and manual verification.

Another possible threat is that the Kaggle solutions might not be
representative. We adopted a two-fold strategy to mitigate that threat.
First, we selected the solutions with the most number of votes and at least
10 votes. Second, we manually verified each program whether it is an
end-to-end DS solution. Since some most voted solutions are only for
introduction and exploratory analysis of the dataset, by manual verification, we excluded those programs. The \gh
projects are also taken from a previously published dataset containing DS
repositories. We further filtered them based on the number of stars and
whether they perform a DS task.

Moreover, since the chosen DS programs from Kaggle and \gh are using Python as the primary language, another question might be on the
generalization of them as DS programs. According to \gh and \sof , Python
has become the most growing language in recent times \cite{gh-survey, so-survey}.
In data science, Python is the most used language because of the availability of
numerous ML, DL and data analysis packages such as Pandas, NumPy, TensorFlow,
Keras, Caffe, Theano, Scikit-Learn and many more.  
\section{Related Work}
\label{sec:related}

Many studies presented ML pipeline in their own context, which can not be generalized for all DS systems. 
\citeauthor{garcia2018context} focused on building an iterative process with three main phases: development, training and inference. They described the interpretation of data and code while integrating the whole lifecycle \cite{garcia2018context}.
\citeauthor{polyzotis2017data} presented the challenges of data management in building production-level ML pipeline in Google around three broad themes: data understanding, data validation and cleaning, and data preparation \cite{polyzotis2017data, polyzotis2018data}. They also provided an overview of an end-to-end large-scale ML pipeline with a data point of view.
Carlton E. Sapp defined ML concepts, business challenges, stages in the
lifecycle, roles of DS teams with comprehensive end-to-end ML
architecture \cite{sapp2017preparing}. This gives us a holistic understanding of the business processes (e.g., acquire, organize, analyze, deliver) of a DS project.

A few other studies try to capture the DS process by surveying and interviewing developers. \citeauthor{roh2018survey} surveyed the data collection techniques in the field of big data. They presented the workflow of data collection answering how to improve data or models in an ML system \cite{roh2018survey}.
Another study identified the software engineering practices and challenges
in building AI applications inside Microsoft development
teams \cite{msml}. They found some key differences in AI software process compared to other domains. They considered a 9-stage workflow for DS software
development.
\citeauthor{hill2016trials} interviewed experienced AI developers and identified problems they face in each stage \cite{hill2016trials}. They also tried to compare the traditional software process and the AI process.
\citeauthor{datanami} presented her own view to build a better ML pipeline \cite{datanami}. They presented three challenges in building ML pipelines: data quality, reliability and accessibility.  

Some articles described ML applications and frameworks which present DS pipelines from industry. For example, 
\textit{Databricks} provides high-level APIs for programming
languages \cite{databrick}.
Team Data Science Process (TDSP) is an agile and iterative process to build
intelligent applications inside Microsoft corporation \cite{microsoftazure}. In a US patent, the authors compared two data analytic lifecycles \cite{todd2017computing}, and presented the difference in the set of parameters with respect to time and cost.
CRoss Industry Standard Process for Data Mining (CRISP-DM) is a 6-stage comprehensive process model for data mining projects across any industry \cite{wirth2000crisp}. 
\citeauthor{mlworkflow} described the workflow of an AI platform \cite{mlworkflow}. They explained tasks completed in each stage with respect to Google Cloud and TensorFlow\cite{abadi2016tensorflow}.
Although there are many papers in the literature presenting DS pipeline, 
there is no comprehensive study that tries to understand and compare DS pipelines in theory and practice.
\section{Conclusion}
\label{sec:conc}

Many software systems today are incorporating a data science pipeline
as their integral part. 
In this work, we argued that to facilitate 
research and practice on data science pipelines, it is essential to understand their nature.  
To that end, we presented a three-pronged comprehensive study of 
data science pipelines in theory, data science pipelines in-the-small,
and data science pipelines in-the-large. 
Our study analyzed three datasets: 
a collection of 71 proposals 
for data science pipelines and related concepts in theory, 
a collection of 105 implementations of data science pipelines from 
Kaggle competitions to understand data science in-the-small, 
and a collection of 21 mature data science projects from GitHub
to understand data science in-the-large.
We have found that DS pipelines differ significantly between these settings. 
Specifically, a number of stages are absent in-the-small, and the DS 
pipelines have a more linear structure. 
The DS pipelines in-the-large have a more complex
structure and feedback loops compared to the theoretical representations. 
We also contribute three representations of DS pipelines that capture 
the essence of our subjects in theory, in-the-small, 
and in-the-large. 

\begin{acks}
	This work was supported in part by US NSF grants CNS-21-20448 and CCF-19-34884. We also thank the reviewers for their insightful comments. All opinions are of the authors and do not reflect the view of sponsors.
\end{acks}

\balance
\bibliographystyle{ACM-Reference-Format}
\bibliography{refs}
 
\appendix

\begin{table*}[t]
	\vspace{14pt}
	\footnotesize
	\setlength\tabcolsep{2pt}
	\caption{Labeled data science pipelines from the subject studies. ACQ: Data acquisition, PRP: Data preparation, STR: Data storage, FTR: Feature engineering, MDL: Modeling, TRN: Training, EVL: Evaluation, PRD: Prediction, INT: Interpretation, CMN: Communication, DPL: Deployment.}
	\label{tab:label}

	Overall goal: \, \greencheckbox \, Describe/propose pipeline, \bluecheckbox  \, Survey/compare/review, \redcheckbox  \, DS optimization, \yellowcheckbox  \, Introduce new method/application \\

	\begin{tabular}{|wc{.7cm}|wl{3.2cm}|wc{.7cm}|wc{.7cm}|wc{.7cm}|wc{.7cm}|wc{.7cm}|wc{.7cm}|wc{.7cm}|wc{.7cm}|wc{.7cm}|wc{.7cm}|wc{.7cm}|wc{.9cm}|wc{.9cm}|wc{.9cm}|}
		\hline
		\rowcolor[HTML]{C0C0C0} 
		\cellcolor[HTML]{C0C0C0}                               & \cellcolor[HTML]{C0C0C0}                                                                               & \multicolumn{3}{c|}{\cellcolor[HTML]{C0C0C0}\textbf{Preprocessing}}                    & \multicolumn{5}{c|}{\cellcolor[HTML]{C0C0C0}\textbf{Modeling}}                                                                                      & \multicolumn{3}{c|}{\cellcolor[HTML]{C0C0C0}\textbf{Post-processing}}                   & \multicolumn{3}{c|}{\cellcolor[HTML]{C0C0C0}\textbf{Involves}}                          \\ \cline{3-16} 
		\rowcolor[HTML]{C0C0C0} 
		\multirow{-2}{*}{\cellcolor[HTML]{C0C0C0}\textbf{Type}} & \multirow{-2}{*}{\cellcolor[HTML]{C0C0C0}\textbf{References}}                                          & \textbf{ACQ}                & \textbf{PRP}                & \textbf{STR}                & \textbf{FTR}                & \textbf{MDL}                & \textbf{TRN}                & \textbf{EVL}                & \textbf{PRD}                & \textbf{INT}                & \textbf{CMN}                & \textbf{DPL}                & \textbf{Cyber}              & \textbf{Physical}           & \textbf{Human}              \\ \hline
		& \citeauthor{olson2016evaluation}, \citeyear{olson2016evaluation} \cite{olson2016evaluation}                         & \redcheckbox                         & \redcheckbox                         & -                           & \redcheckbox                         & \redcheckbox                         & \redcheckbox                         & \redcheckbox                         & \redcheckbox                         & -                           & -                           & -                           & \redcheckbox                         & -                           & -                           \\
		& \cellcolor[HTML]{EFEFEF}\citeauthor{miao2017ds}, \citeyear{miao2017ds} \cite{miao2017ds}                            & \cellcolor[HTML]{EFEFEF}\greencheckbox & \cellcolor[HTML]{EFEFEF}-   & \cellcolor[HTML]{EFEFEF}-   & \cellcolor[HTML]{EFEFEF}-   & \cellcolor[HTML]{EFEFEF}\greencheckbox & \cellcolor[HTML]{EFEFEF}\greencheckbox & \cellcolor[HTML]{EFEFEF}\greencheckbox & \cellcolor[HTML]{EFEFEF}\greencheckbox & \cellcolor[HTML]{EFEFEF}-   & \cellcolor[HTML]{EFEFEF}-   & \cellcolor[HTML]{EFEFEF}\greencheckbox & \cellcolor[HTML]{EFEFEF}\greencheckbox & \cellcolor[HTML]{EFEFEF}-   & \cellcolor[HTML]{EFEFEF}-   \\
		& \citeauthor{garcia2018context}, \citeyear{garcia2018context} \cite{garcia2018context}                               & -                           & \greencheckbox                         & -                           & -                           & \greencheckbox                         & \greencheckbox                         & \greencheckbox                         & \greencheckbox                         & -                           & \greencheckbox                         & -                           & \greencheckbox                         & -                           & -                           \\
		& \cellcolor[HTML]{EFEFEF}\citeauthor{databrick}, \citeyear{databrick} \cite{databrick}                               & \cellcolor[HTML]{EFEFEF}\greencheckbox & \cellcolor[HTML]{EFEFEF}\greencheckbox & \cellcolor[HTML]{EFEFEF}-   & \cellcolor[HTML]{EFEFEF}\greencheckbox & \cellcolor[HTML]{EFEFEF}\greencheckbox & \cellcolor[HTML]{EFEFEF}-   & \cellcolor[HTML]{EFEFEF}-   & \cellcolor[HTML]{EFEFEF}-   & \cellcolor[HTML]{EFEFEF}-   & \cellcolor[HTML]{EFEFEF}-   & \cellcolor[HTML]{EFEFEF}-   & \cellcolor[HTML]{EFEFEF}\greencheckbox & \cellcolor[HTML]{EFEFEF}-   & \cellcolor[HTML]{EFEFEF}-   \\
		& \citeauthor{microsoft}, \citeyear{microsoft} \cite{microsoft}                                                       & -                           & \greencheckbox                         & \greencheckbox                         & \greencheckbox                         & \greencheckbox                         & \greencheckbox                         & \greencheckbox                         & \greencheckbox                         & -                           & -                           & \greencheckbox                         & \greencheckbox                         & -                           & -                           \\
		& \cellcolor[HTML]{EFEFEF}\citeauthor{datanami}, \citeyear{datanami} \cite{datanami}                                  & \cellcolor[HTML]{EFEFEF}-   & \cellcolor[HTML]{EFEFEF}\greencheckbox & \cellcolor[HTML]{EFEFEF}-   & \cellcolor[HTML]{EFEFEF}\greencheckbox & \cellcolor[HTML]{EFEFEF}\greencheckbox & \cellcolor[HTML]{EFEFEF}\greencheckbox & \cellcolor[HTML]{EFEFEF}-   & \cellcolor[HTML]{EFEFEF}\greencheckbox & \cellcolor[HTML]{EFEFEF}-   & \cellcolor[HTML]{EFEFEF}-   & \cellcolor[HTML]{EFEFEF}-   & \cellcolor[HTML]{EFEFEF}\greencheckbox & \cellcolor[HTML]{EFEFEF}-   & \cellcolor[HTML]{EFEFEF}-   \\
		& \citeauthor{towardsdatascience}, \citeyear{towardsdatascience} \cite{towardsdatascience}                            & -                           & \greencheckbox                         & -                           & -                           & \greencheckbox                         & \greencheckbox                         & -                           & -                           & -                           & -                           & -                           & \greencheckbox                         & -                           & -                           \\
		& \cellcolor[HTML]{EFEFEF}\citeauthor{polyzotis2018data}, \citeyear{polyzotis2018data} \cite{polyzotis2018data}       & \cellcolor[HTML]{EFEFEF}-   & \cellcolor[HTML]{EFEFEF}\bluecheckbox & \cellcolor[HTML]{EFEFEF}-   & \cellcolor[HTML]{EFEFEF}\bluecheckbox & \cellcolor[HTML]{EFEFEF}\bluecheckbox & \cellcolor[HTML]{EFEFEF}\bluecheckbox & \cellcolor[HTML]{EFEFEF}\bluecheckbox & \cellcolor[HTML]{EFEFEF}-   & \cellcolor[HTML]{EFEFEF}-   & \cellcolor[HTML]{EFEFEF}-   & \cellcolor[HTML]{EFEFEF}\bluecheckbox & \cellcolor[HTML]{EFEFEF}\bluecheckbox & \cellcolor[HTML]{EFEFEF}-   & \cellcolor[HTML]{EFEFEF}-   \\
		& \citeauthor{roh2018survey}, \citeyear{roh2018survey} \cite{roh2018survey}                                           & \bluecheckbox                         & \bluecheckbox                         & -                           & \bluecheckbox                         & \bluecheckbox                         & -                           & -                           & -                           & -                           & \bluecheckbox                         & -                           & \bluecheckbox                         & \bluecheckbox                         & -                           \\
		& \cellcolor[HTML]{EFEFEF}\citeauthor{miao2017provdb}, \citeyear{miao2017provdb} \cite{miao2017provdb}                & \cellcolor[HTML]{EFEFEF}\redcheckbox & \cellcolor[HTML]{EFEFEF}-   & \cellcolor[HTML]{EFEFEF}\redcheckbox & \cellcolor[HTML]{EFEFEF}-   & \cellcolor[HTML]{EFEFEF}\redcheckbox & \cellcolor[HTML]{EFEFEF}-   & \cellcolor[HTML]{EFEFEF}-   & \cellcolor[HTML]{EFEFEF}-   & \cellcolor[HTML]{EFEFEF}-   & \cellcolor[HTML]{EFEFEF}-   & \cellcolor[HTML]{EFEFEF}-   & \cellcolor[HTML]{EFEFEF}\redcheckbox & \cellcolor[HTML]{EFEFEF}-   & \cellcolor[HTML]{EFEFEF}-   \\
		& \citeauthor{sparks2017keystoneml}, \citeyear{sparks2017keystoneml} \cite{sparks2017keystoneml}                      & -                           & \redcheckbox                         & -                           & \redcheckbox                         & \redcheckbox                         & \redcheckbox                         & -                           & -                           & -                           & -                           & -                           & \redcheckbox                         & -                           & -                           \\
		& \cellcolor[HTML]{EFEFEF}\citeauthor{mlsteps}, \citeyear{mlsteps} \cite{mlsteps}                                     & \cellcolor[HTML]{EFEFEF}\greencheckbox & \cellcolor[HTML]{EFEFEF}\greencheckbox & \cellcolor[HTML]{EFEFEF}-   & \cellcolor[HTML]{EFEFEF}\greencheckbox & \cellcolor[HTML]{EFEFEF}\greencheckbox & \cellcolor[HTML]{EFEFEF}\greencheckbox & \cellcolor[HTML]{EFEFEF}\greencheckbox & \cellcolor[HTML]{EFEFEF}\greencheckbox & \cellcolor[HTML]{EFEFEF}-   & \cellcolor[HTML]{EFEFEF}-   & \cellcolor[HTML]{EFEFEF}-   & \cellcolor[HTML]{EFEFEF}\greencheckbox & \cellcolor[HTML]{EFEFEF}-   & \cellcolor[HTML]{EFEFEF}-   \\
		& \citeauthor{baylor2017tfx}, \citeyear{baylor2017tfx} \cite{baylor2017tfx}                                           & -                           & \yellowcheckbox                         & -                           & \yellowcheckbox                         & \yellowcheckbox                         & \yellowcheckbox                         & \yellowcheckbox                         & -                           & -                           & \yellowcheckbox                         & \yellowcheckbox                         & \yellowcheckbox                         & -                           & -                           \\
		& \cellcolor[HTML]{EFEFEF}\citeauthor{abadi2016tensorflow}, \citeyear{abadi2016tensorflow} \cite{abadi2016tensorflow} & \cellcolor[HTML]{EFEFEF}-   & \cellcolor[HTML]{EFEFEF}\yellowcheckbox & \cellcolor[HTML]{EFEFEF}\yellowcheckbox & \cellcolor[HTML]{EFEFEF}-   & \cellcolor[HTML]{EFEFEF}\yellowcheckbox & \cellcolor[HTML]{EFEFEF}\yellowcheckbox & \cellcolor[HTML]{EFEFEF}-   & \cellcolor[HTML]{EFEFEF}-   & \cellcolor[HTML]{EFEFEF}-   & \cellcolor[HTML]{EFEFEF}-   & \cellcolor[HTML]{EFEFEF}-   & \cellcolor[HTML]{EFEFEF}\yellowcheckbox & \cellcolor[HTML]{EFEFEF}-   & \cellcolor[HTML]{EFEFEF}-   \\
		& \citeauthor{chilimbi2014project}, \citeyear{chilimbi2014project} \cite{chilimbi2014project}                         & -                           & -                           & -                           & \yellowcheckbox                         & \yellowcheckbox                         & \yellowcheckbox                         & -                           & \yellowcheckbox                         & -                           & -                           & -                           & \yellowcheckbox                         & -                           & -                           \\
		& \cellcolor[HTML]{EFEFEF}\citeauthor{kraska2013mlbase}, \citeyear{kraska2013mlbase} \cite{kraska2013mlbase}          & \cellcolor[HTML]{EFEFEF}-   & \cellcolor[HTML]{EFEFEF}-   & \cellcolor[HTML]{EFEFEF}\yellowcheckbox & \cellcolor[HTML]{EFEFEF}\yellowcheckbox & \cellcolor[HTML]{EFEFEF}\yellowcheckbox & \cellcolor[HTML]{EFEFEF}\yellowcheckbox & \cellcolor[HTML]{EFEFEF}\yellowcheckbox & \cellcolor[HTML]{EFEFEF}-   & \cellcolor[HTML]{EFEFEF}\yellowcheckbox & \cellcolor[HTML]{EFEFEF}-   & \cellcolor[HTML]{EFEFEF}-   & \cellcolor[HTML]{EFEFEF}\yellowcheckbox & \cellcolor[HTML]{EFEFEF}-   & \cellcolor[HTML]{EFEFEF}-   \\
		& \citeauthor{sculley2015hidden}, \citeyear{sculley2015hidden} \cite{sculley2015hidden}                               & \bluecheckbox                         & -                           & -                           & \bluecheckbox                         & \bluecheckbox                         & -                           & -                           & -                           & -                           & -                           & \bluecheckbox                         & \bluecheckbox                         & -                           & -                           \\
		& \cellcolor[HTML]{EFEFEF}\citeauthor{mlmodeling}, \citeyear{mlmodeling} \cite{mlmodeling}                            & \cellcolor[HTML]{EFEFEF}\greencheckbox & \cellcolor[HTML]{EFEFEF}\greencheckbox & \cellcolor[HTML]{EFEFEF}-   & \cellcolor[HTML]{EFEFEF}\greencheckbox & \cellcolor[HTML]{EFEFEF}\greencheckbox & \cellcolor[HTML]{EFEFEF}\greencheckbox & \cellcolor[HTML]{EFEFEF}\greencheckbox & \cellcolor[HTML]{EFEFEF}\greencheckbox & \cellcolor[HTML]{EFEFEF}-   & \cellcolor[HTML]{EFEFEF}-   & \cellcolor[HTML]{EFEFEF}\greencheckbox & \cellcolor[HTML]{EFEFEF}\greencheckbox & \cellcolor[HTML]{EFEFEF}-   & \cellcolor[HTML]{EFEFEF}-   \\
		& \citeauthor{mlworkflow}, \citeyear{mlworkflow} \cite{mlworkflow}                                                          & \greencheckbox                         & \greencheckbox                         & -                           & -                           & \greencheckbox                         & \greencheckbox                         & \greencheckbox                         & \greencheckbox                         & -                           & -                           & -                           & \greencheckbox                         & -                           & -                           \\
		& \cellcolor[HTML]{EFEFEF}\citeauthor{msml}, \citeyear{msml} \cite{msml}                                              & \cellcolor[HTML]{EFEFEF}\bluecheckbox & \cellcolor[HTML]{EFEFEF}\bluecheckbox & \cellcolor[HTML]{EFEFEF}-   & \cellcolor[HTML]{EFEFEF}\bluecheckbox & \cellcolor[HTML]{EFEFEF}\bluecheckbox & \cellcolor[HTML]{EFEFEF}\bluecheckbox & \cellcolor[HTML]{EFEFEF}\bluecheckbox & \cellcolor[HTML]{EFEFEF}\bluecheckbox & \cellcolor[HTML]{EFEFEF}-   & \cellcolor[HTML]{EFEFEF}-   & \cellcolor[HTML]{EFEFEF}\bluecheckbox & \cellcolor[HTML]{EFEFEF}\bluecheckbox & \cellcolor[HTML]{EFEFEF}-   & \cellcolor[HTML]{EFEFEF}-   \\
		& \citeauthor{van2017versioning}, \citeyear{van2017versioning} \cite{van2017versioning}                               & \redcheckbox                         & \redcheckbox                         & -                           & \redcheckbox                         & -                           & -                           & -                           & \redcheckbox                         & -                           & -                           & -                           & \redcheckbox                         & -                           & \redcheckbox                         \\
		& \cellcolor[HTML]{EFEFEF}\citeauthor{hill2016trials}, \citeyear{hill2016trials} \cite{hill2016trials}                & \cellcolor[HTML]{EFEFEF}\greencheckbox & \cellcolor[HTML]{EFEFEF}-   & \cellcolor[HTML]{EFEFEF}-   & \cellcolor[HTML]{EFEFEF}\greencheckbox & \cellcolor[HTML]{EFEFEF}\greencheckbox & \cellcolor[HTML]{EFEFEF}-   & \cellcolor[HTML]{EFEFEF}\greencheckbox & \cellcolor[HTML]{EFEFEF}-   & \cellcolor[HTML]{EFEFEF}-   & \cellcolor[HTML]{EFEFEF}-   & \cellcolor[HTML]{EFEFEF}-   & \cellcolor[HTML]{EFEFEF}\greencheckbox & \cellcolor[HTML]{EFEFEF}-   & \cellcolor[HTML]{EFEFEF}-   \\
		& \citeauthor{shang2019democratizing}, \citeyear{shang2019democratizing} \cite{shang2019democratizing}                & -                           & \redcheckbox                         & -                           & \redcheckbox                         & \redcheckbox                         & -                           & -                           & -                           & -                           & -                           & -                           & \redcheckbox                         & -                           & -                           \\                                 
		& \cellcolor[HTML]{EFEFEF}\citeauthor{zhang2016flash}, \citeyear{zhang2016flash} \cite{zhang2016flash}                & \cellcolor[HTML]{EFEFEF}-   & \cellcolor[HTML]{EFEFEF}\redcheckbox & \cellcolor[HTML]{EFEFEF}-   & \cellcolor[HTML]{EFEFEF}\redcheckbox & \cellcolor[HTML]{EFEFEF}\redcheckbox & \cellcolor[HTML]{EFEFEF}-   & \cellcolor[HTML]{EFEFEF}-   & \cellcolor[HTML]{EFEFEF}-   & \cellcolor[HTML]{EFEFEF}-   & \cellcolor[HTML]{EFEFEF}-   & \cellcolor[HTML]{EFEFEF}-   & \cellcolor[HTML]{EFEFEF}\redcheckbox & \cellcolor[HTML]{EFEFEF}-   & \cellcolor[HTML]{EFEFEF}-   \\
		& \citeauthor{gil2018p4ml}, \citeyear{gil2018p4ml} \cite{gil2018p4ml}                         & -                         & \greencheckbox                         & -                           & \greencheckbox                         & \greencheckbox                         & -                         & \greencheckbox                         & \greencheckbox                         & -                           & -                           & -                           & \greencheckbox                         & -                           & -                           \\
		& \cellcolor[HTML]{EFEFEF}\citeauthor{sadiq2018data}, \citeyear{sadiq2018data} \cite{sadiq2018data}                         & \cellcolor[HTML]{EFEFEF}\greencheckbox                         & \cellcolor[HTML]{EFEFEF}\greencheckbox                         & \cellcolor[HTML]{EFEFEF}\greencheckbox                           & \cellcolor[HTML]{EFEFEF}\greencheckbox                         & \cellcolor[HTML]{EFEFEF}\greencheckbox                         & \cellcolor[HTML]{EFEFEF}-                         & \cellcolor[HTML]{EFEFEF}\greencheckbox                         & \cellcolor[HTML]{EFEFEF}-                         & \cellcolor[HTML]{EFEFEF}\greencheckbox                           & \cellcolor[HTML]{EFEFEF}-                           & \cellcolor[HTML]{EFEFEF}-                           & \cellcolor[HTML]{EFEFEF}\greencheckbox                         & \cellcolor[HTML]{EFEFEF}-                           & \cellcolor[HTML]{EFEFEF}-                           \\
		& \citeauthor{zhou2020semfe}, \citeyear{zhou2020semfe} \cite{zhou2020semfe}                         & \greencheckbox                         & \greencheckbox                         & -                           & -                         & \greencheckbox                         & -                         & -                         & \greencheckbox                         & \greencheckbox                           & -                           & \greencheckbox                           & \greencheckbox                         & \greencheckbox                           & -                           \\
		& \cellcolor[HTML]{EFEFEF}\citeauthor{aggarwal2019can}, \citeyear{aggarwal2019can} \cite{aggarwal2019can}                         & \cellcolor[HTML]{EFEFEF}-                         & \cellcolor[HTML]{EFEFEF}\greencheckbox                         & \cellcolor[HTML]{EFEFEF}-                           & \cellcolor[HTML]{EFEFEF}\greencheckbox                         & \cellcolor[HTML]{EFEFEF}\greencheckbox                         & \cellcolor[HTML]{EFEFEF}\greencheckbox                         & \cellcolor[HTML]{EFEFEF}-                         & \cellcolor[HTML]{EFEFEF}\greencheckbox                         & \cellcolor[HTML]{EFEFEF}-                           & \cellcolor[HTML]{EFEFEF}-                           & \cellcolor[HTML]{EFEFEF}-                           & \cellcolor[HTML]{EFEFEF}\greencheckbox                         & \cellcolor[HTML]{EFEFEF}-                           & \cellcolor[HTML]{EFEFEF}-                           \\
		& \citeauthor{toreini2020relationship}, \citeyear{toreini2020relationship} \cite{toreini2020relationship}                         & \greencheckbox                         & \greencheckbox                         & -                           & \greencheckbox                         & -                         & \greencheckbox                         & \greencheckbox                         & \greencheckbox                         & -                           & -                           & -                           & \greencheckbox                         & -                           & -                           \\
		& \cellcolor[HTML]{EFEFEF}\citeauthor{ashmore2019assuring}, \citeyear{ashmore2019assuring} \cite{ashmore2019assuring}                         & \cellcolor[HTML]{EFEFEF}\bluecheckbox                         & \cellcolor[HTML]{EFEFEF}\bluecheckbox                         & \cellcolor[HTML]{EFEFEF}\bluecheckbox                           & \cellcolor[HTML]{EFEFEF}\bluecheckbox                         & \cellcolor[HTML]{EFEFEF}\bluecheckbox                         & \cellcolor[HTML]{EFEFEF}\bluecheckbox                         & \cellcolor[HTML]{EFEFEF}\bluecheckbox                         & \cellcolor[HTML]{EFEFEF}\bluecheckbox                         & \cellcolor[HTML]{EFEFEF}\bluecheckbox                           & \cellcolor[HTML]{EFEFEF}\bluecheckbox                           & \cellcolor[HTML]{EFEFEF}\bluecheckbox                           & \cellcolor[HTML]{EFEFEF}\bluecheckbox                         & \cellcolor[HTML]{EFEFEF}\bluecheckbox                           & \cellcolor[HTML]{EFEFEF}\bluecheckbox                           \\
		& \citeauthor{shashanka2019pipe}, \citeyear{shashanka2019pipe} \cite{shashanka2019pipe}                         & \greencheckbox                         & \greencheckbox                         & -                           & \greencheckbox                         & \greencheckbox                         & \greencheckbox                         & \greencheckbox                         & \greencheckbox                         & \greencheckbox                           & --                           & --                           & \greencheckbox                         & --                           & --                           \\
		& \cellcolor[HTML]{EFEFEF}\citeauthor{valohai2020pipeline}, \citeyear{valohai2020pipeline} \cite{valohai2020pipeline}                         & \cellcolor[HTML]{EFEFEF}\greencheckbox                         & \cellcolor[HTML]{EFEFEF}\greencheckbox                         & \cellcolor[HTML]{EFEFEF}\greencheckbox                           & \cellcolor[HTML]{EFEFEF}-                         & \cellcolor[HTML]{EFEFEF}\greencheckbox                         & \cellcolor[HTML]{EFEFEF}\greencheckbox                         & \cellcolor[HTML]{EFEFEF}\greencheckbox                         & \cellcolor[HTML]{EFEFEF}-                         & \cellcolor[HTML]{EFEFEF}\greencheckbox                           & \cellcolor[HTML]{EFEFEF}--                           & \cellcolor[HTML]{EFEFEF}--                           & \cellcolor[HTML]{EFEFEF}\greencheckbox                         & \cellcolor[HTML]{EFEFEF}--                           & \cellcolor[HTML]{EFEFEF}--                           \\
		\multirow{-35}{*}{\rotatebox{90}{Machine learning process}}
		& \citeauthor{daum2016pipeline}, \citeyear{daum2016pipeline} \cite{daum2016pipeline}                         & \yellowcheckbox                         & \yellowcheckbox                         & -                           & -                         & \yellowcheckbox                         & \yellowcheckbox                         & \yellowcheckbox                         & \yellowcheckbox                         & -                           & -                           & -                           & \yellowcheckbox                         & -                           & -                           \\
		\hline
		\hline
		& \citeauthor{todd2017computing}, \citeyear{todd2017computing} \cite{todd2017computing}                               & \yellowcheckbox                         & \yellowcheckbox                         & \yellowcheckbox                         & \yellowcheckbox                         & \yellowcheckbox                         & -                           & -                           & \yellowcheckbox                         & -                           & \yellowcheckbox                         & -                           & \yellowcheckbox                         & -                           & \yellowcheckbox                         \\
		& \cellcolor[HTML]{EFEFEF}\citeauthor{zhang2017framework}, \citeyear{zhang2017framework} \cite{zhang2017framework}    & \cellcolor[HTML]{EFEFEF}\greencheckbox & \cellcolor[HTML]{EFEFEF}\greencheckbox & \cellcolor[HTML]{EFEFEF}\greencheckbox & \cellcolor[HTML]{EFEFEF}-   & \cellcolor[HTML]{EFEFEF}\greencheckbox & \cellcolor[HTML]{EFEFEF}-   & \cellcolor[HTML]{EFEFEF}-   & \cellcolor[HTML]{EFEFEF}\greencheckbox & \cellcolor[HTML]{EFEFEF}-   & \cellcolor[HTML]{EFEFEF}\greencheckbox & \cellcolor[HTML]{EFEFEF}-   & \cellcolor[HTML]{EFEFEF}\greencheckbox & \cellcolor[HTML]{EFEFEF}\greencheckbox & \cellcolor[HTML]{EFEFEF}-   \\
		& \citeauthor{sapp2017preparing}, \citeyear{sapp2017preparing} \cite{sapp2017preparing}                               & \bluecheckbox                         & \bluecheckbox                         & \bluecheckbox                         & \bluecheckbox                         & \bluecheckbox                         & \bluecheckbox                         & \bluecheckbox                         & \bluecheckbox                         & -                           & \bluecheckbox                         & \bluecheckbox                         & \bluecheckbox                         & \bluecheckbox                         & \bluecheckbox                         \\
		& \cellcolor[HTML]{EFEFEF}\citeauthor{landset2015survey}, \citeyear{landset2015survey} \cite{landset2015survey}       & \cellcolor[HTML]{EFEFEF}-   & \cellcolor[HTML]{EFEFEF}-   & \cellcolor[HTML]{EFEFEF}\redcheckbox & \cellcolor[HTML]{EFEFEF}-   & \cellcolor[HTML]{EFEFEF}\redcheckbox & \cellcolor[HTML]{EFEFEF}\redcheckbox & \cellcolor[HTML]{EFEFEF}\redcheckbox & \cellcolor[HTML]{EFEFEF}\redcheckbox & \cellcolor[HTML]{EFEFEF}-   & \cellcolor[HTML]{EFEFEF}-   & \cellcolor[HTML]{EFEFEF}\redcheckbox & \cellcolor[HTML]{EFEFEF}\redcheckbox & \cellcolor[HTML]{EFEFEF}-   & \cellcolor[HTML]{EFEFEF}-   \\
		& \citeauthor{polyzotis2017data}, \citeyear{polyzotis2017data} \cite{polyzotis2017data}                               & -                           & -                           & -                           & -                           & \bluecheckbox                         & \bluecheckbox                         & -                           & \bluecheckbox                         & -                           & -                           & \bluecheckbox                         & \bluecheckbox                         & -                           & -                           \\
		& \cellcolor[HTML]{EFEFEF}\citeauthor{hu2014toward}, \citeyear{hu2014toward} \cite{hu2014toward}                      & \cellcolor[HTML]{EFEFEF}\bluecheckbox & \cellcolor[HTML]{EFEFEF}\bluecheckbox & \cellcolor[HTML]{EFEFEF}\bluecheckbox & \cellcolor[HTML]{EFEFEF}-   & \cellcolor[HTML]{EFEFEF}-   & \cellcolor[HTML]{EFEFEF}-   & \cellcolor[HTML]{EFEFEF}-   & \cellcolor[HTML]{EFEFEF}-   & \cellcolor[HTML]{EFEFEF}-   & \cellcolor[HTML]{EFEFEF}-   & \cellcolor[HTML]{EFEFEF}-   & \cellcolor[HTML]{EFEFEF}\bluecheckbox & \cellcolor[HTML]{EFEFEF}\bluecheckbox & \cellcolor[HTML]{EFEFEF}\bluecheckbox \\
		& \citeauthor{demchenko2012addressing}, \citeyear{demchenko2012addressing} \cite{demchenko2012addressing}             & \greencheckbox                         & \greencheckbox                         & \greencheckbox                         & -                           & -                           & -                           & -                           & -                           & -                           & \greencheckbox                         & -                           & \greencheckbox                         & -                           & -                           \\
		& \cellcolor[HTML]{EFEFEF}\citeauthor{khan2017survey}, \citeyear{khan2017survey} \cite{khan2017survey}                      & \cellcolor[HTML]{EFEFEF}- & \cellcolor[HTML]{EFEFEF}\bluecheckbox & \cellcolor[HTML]{EFEFEF}- & \cellcolor[HTML]{EFEFEF}\bluecheckbox   & \cellcolor[HTML]{EFEFEF}\bluecheckbox   & \cellcolor[HTML]{EFEFEF}-   & \cellcolor[HTML]{EFEFEF}\bluecheckbox   & \cellcolor[HTML]{EFEFEF}\bluecheckbox   & \cellcolor[HTML]{EFEFEF}\bluecheckbox   & \cellcolor[HTML]{EFEFEF}\bluecheckbox   & \cellcolor[HTML]{EFEFEF}-   & \cellcolor[HTML]{EFEFEF}\bluecheckbox & \cellcolor[HTML]{EFEFEF}- & \cellcolor[HTML]{EFEFEF}- \\
		& \citeauthor{el2018data}, \citeyear{el2018data} \cite{el2018data}             & \greencheckbox                         & \greencheckbox                         & \greencheckbox                         & \greencheckbox                           & -                           & -                           & \greencheckbox                           & -                           & --                           & --                         & \greencheckbox                           & \greencheckbox                         & -                           & -                           \\
		& \cellcolor[HTML]{EFEFEF}\citeauthor{hummer2019modelops}, \citeyear{hummer2019modelops} \cite{hummer2019modelops}                      & \cellcolor[HTML]{EFEFEF}- & \cellcolor[HTML]{EFEFEF}\yellowcheckbox & \cellcolor[HTML]{EFEFEF}- & \cellcolor[HTML]{EFEFEF}-   & \cellcolor[HTML]{EFEFEF}\yellowcheckbox   & \cellcolor[HTML]{EFEFEF}\yellowcheckbox   & \cellcolor[HTML]{EFEFEF}\yellowcheckbox   & \cellcolor[HTML]{EFEFEF}\yellowcheckbox   & \cellcolor[HTML]{EFEFEF}-   & \cellcolor[HTML]{EFEFEF}\yellowcheckbox   & \cellcolor[HTML]{EFEFEF}\yellowcheckbox   & \cellcolor[HTML]{EFEFEF}\yellowcheckbox & \cellcolor[HTML]{EFEFEF}- & \cellcolor[HTML]{EFEFEF}\yellowcheckbox \\
		& \citeauthor{mehmet2020pipe}, \citeyear{mehmet2020pipe} \cite{mehmet2020pipe}             & \greencheckbox                         & \greencheckbox                         & \greencheckbox                         & \greencheckbox                           & \greencheckbox                           & -                           & \greencheckbox                           & -                           & \greencheckbox                           & \greencheckbox                         & \greencheckbox                           & \greencheckbox                         & -                           & -                           \\
		& \cellcolor[HTML]{EFEFEF}\citeauthor{stephanie2019}, \citeyear{stephanie2019} \cite{stephanie2019}                      & \cellcolor[HTML]{EFEFEF}\greencheckbox & \cellcolor[HTML]{EFEFEF}\greencheckbox & \cellcolor[HTML]{EFEFEF}- & \cellcolor[HTML]{EFEFEF}-   & \cellcolor[HTML]{EFEFEF}\greencheckbox   & \cellcolor[HTML]{EFEFEF}\greencheckbox   & \cellcolor[HTML]{EFEFEF}\greencheckbox   & \cellcolor[HTML]{EFEFEF}-   & \cellcolor[HTML]{EFEFEF}-   & \cellcolor[HTML]{EFEFEF}\greencheckbox   & \cellcolor[HTML]{EFEFEF}-   & \cellcolor[HTML]{EFEFEF}\greencheckbox & \cellcolor[HTML]{EFEFEF}- & \cellcolor[HTML]{EFEFEF}- \\
		\multirow{-14}{*}{\rotatebox{90}{Big data management}}                                  
		& \citeauthor{tim2018}, \citeyear{tim2018} \cite{tim2018}             & -                         & \greencheckbox                         & -                         & \greencheckbox                           & \greencheckbox                           & \greencheckbox                           & \greencheckbox                           & \greencheckbox                           & -                           & --                         & \greencheckbox                           & \greencheckbox                         & -                           & -                           \\
				
		\hline
\end{tabular}
\end{table*} 
\begin{table*}[t]
	\vspace{14pt}
	\footnotesize
	\setlength\tabcolsep{2pt}

	\begin{tabular}{|wc{.7cm}|wl{3.2cm}|wc{.7cm}|wc{.7cm}|wc{.7cm}|wc{.7cm}|wc{.7cm}|wc{.7cm}|wc{.7cm}|wc{.7cm}|wc{.7cm}|wc{.7cm}|wc{.7cm}|wc{.9cm}|wc{.9cm}|wc{.9cm}|}
		\hline
		\rowcolor[HTML]{C0C0C0} 
		\cellcolor[HTML]{C0C0C0}                               & \cellcolor[HTML]{C0C0C0}                                                                               & \multicolumn{3}{c|}{\cellcolor[HTML]{C0C0C0}\textbf{Preprocessing}}                    & \multicolumn{5}{c|}{\cellcolor[HTML]{C0C0C0}\textbf{Modeling}}                                                                                      & \multicolumn{3}{c|}{\cellcolor[HTML]{C0C0C0}\textbf{Post-processing}}                   & \multicolumn{3}{c|}{\cellcolor[HTML]{C0C0C0}\textbf{Involves}}                          \\ \cline{3-16} 
		\rowcolor[HTML]{C0C0C0} 
		\multirow{-2}{*}{\cellcolor[HTML]{C0C0C0}\textbf{Type}} & \multirow{-2}{*}{\cellcolor[HTML]{C0C0C0}\textbf{References}}                                          & \textbf{ACQ}                & \textbf{PRP}                & \textbf{STR}                & \textbf{FTR}                & \textbf{MDL}                & \textbf{TRN}                & \textbf{EVL}                & \textbf{PRD}                & \textbf{INT}                & \textbf{CMN}                & \textbf{DPL}                & \textbf{Cyber}              & \textbf{Physical}           & \textbf{Human}              \\ \hline
		& \cellcolor[HTML]{EFEFEF}\citeauthor{pouchard2016revisiting}, \citeyear{pouchard2016revisiting} \cite{pouchard2016revisiting}                & \cellcolor[HTML]{EFEFEF}\greencheckbox & \cellcolor[HTML]{EFEFEF}\greencheckbox & \cellcolor[HTML]{EFEFEF}\greencheckbox & \cellcolor[HTML]{EFEFEF}- & \cellcolor[HTML]{EFEFEF}- & \cellcolor[HTML]{EFEFEF}- & \cellcolor[HTML]{EFEFEF}- & \cellcolor[HTML]{EFEFEF}- & \cellcolor[HTML]{EFEFEF}- & \cellcolor[HTML]{EFEFEF}\greencheckbox & \cellcolor[HTML]{EFEFEF}- & \cellcolor[HTML]{EFEFEF}\greencheckbox & \cellcolor[HTML]{EFEFEF}- & \cellcolor[HTML]{EFEFEF}\greencheckbox \\
		
		& \citeauthor{microsoftazure}, \citeyear{microsoftazure} \cite{microsoftazure}                & \greencheckbox & \greencheckbox & \greencheckbox & \greencheckbox & \greencheckbox & \greencheckbox & \greencheckbox & \greencheckbox & -   & \greencheckbox & \greencheckbox & \greencheckbox & -   & \greencheckbox \\
		
		& \cellcolor[HTML]{EFEFEF}\citeauthor{berman2018realizing}, \citeyear{berman2018realizing} \cite{berman2018realizing}                & \cellcolor[HTML]{EFEFEF}\bluecheckbox & \cellcolor[HTML]{EFEFEF}\bluecheckbox & \cellcolor[HTML]{EFEFEF}\bluecheckbox & \cellcolor[HTML]{EFEFEF}- & \cellcolor[HTML]{EFEFEF}\bluecheckbox & \cellcolor[HTML]{EFEFEF}- & \cellcolor[HTML]{EFEFEF}- & \cellcolor[HTML]{EFEFEF}- & \cellcolor[HTML]{EFEFEF}- & \cellcolor[HTML]{EFEFEF}\bluecheckbox & \cellcolor[HTML]{EFEFEF}\bluecheckbox & \cellcolor[HTML]{EFEFEF}\bluecheckbox & \cellcolor[HTML]{EFEFEF}\bluecheckbox & \cellcolor[HTML]{EFEFEF}\bluecheckbox \\
		
		& \citeauthor{dsl-sudeep}, \citeyear{dsl-sudeep} \cite{dsl-sudeep}                                                    & \greencheckbox                         & \greencheckbox                         & \greencheckbox                         & \greencheckbox                         & \greencheckbox                         & \greencheckbox                         & \greencheckbox                         & \greencheckbox                         & -                           & \greencheckbox                         & -                           & \greencheckbox                         & -                           & \greencheckbox                         \\
		
		& \cellcolor[HTML]{EFEFEF}\citeauthor{nguyen2019machine}, \citeyear{nguyen2019machine} \cite{nguyen2019machine}                & \cellcolor[HTML]{EFEFEF}\bluecheckbox & \cellcolor[HTML]{EFEFEF}\bluecheckbox & \cellcolor[HTML]{EFEFEF}- & \cellcolor[HTML]{EFEFEF}\bluecheckbox & \cellcolor[HTML]{EFEFEF}\bluecheckbox & \cellcolor[HTML]{EFEFEF}\bluecheckbox & \cellcolor[HTML]{EFEFEF}\bluecheckbox & \cellcolor[HTML]{EFEFEF}\bluecheckbox & \cellcolor[HTML]{EFEFEF}- & \cellcolor[HTML]{EFEFEF}- & \cellcolor[HTML]{EFEFEF}- & \cellcolor[HTML]{EFEFEF}\bluecheckbox & \cellcolor[HTML]{EFEFEF}- & \cellcolor[HTML]{EFEFEF}- \\
		
		& \citeauthor{ruegg2014completing}, \citeyear{ruegg2014completing} \cite{ruegg2014completing}                & \greencheckbox & \greencheckbox & - & - & - & - & - & - & \greencheckbox & \greencheckbox & - & \greencheckbox & - & - \\
		
		& \cellcolor[HTML]{EFEFEF}\citeauthor{gandomi2015beyond}, \citeyear{gandomi2015beyond} \cite{gandomi2015beyond}                & \cellcolor[HTML]{EFEFEF}\bluecheckbox & \cellcolor[HTML]{EFEFEF}\bluecheckbox & \cellcolor[HTML]{EFEFEF}\bluecheckbox & \cellcolor[HTML]{EFEFEF}\bluecheckbox & \cellcolor[HTML]{EFEFEF}\bluecheckbox & \cellcolor[HTML]{EFEFEF}- & \cellcolor[HTML]{EFEFEF}\bluecheckbox & \cellcolor[HTML]{EFEFEF}- & \cellcolor[HTML]{EFEFEF}\bluecheckbox & \cellcolor[HTML]{EFEFEF}- & \cellcolor[HTML]{EFEFEF}- & \cellcolor[HTML]{EFEFEF}\bluecheckbox & \cellcolor[HTML]{EFEFEF}- & \cellcolor[HTML]{EFEFEF}- \\
		
		& \citeauthor{ball2012review}, \citeyear{ball2012review} \cite{ball2012review}                & \bluecheckbox & - & \bluecheckbox & - & - & - & \bluecheckbox & - & - & \bluecheckbox & - & \bluecheckbox & - & - \\
		
		& \cellcolor[HTML]{EFEFEF}\citeauthor{wing2019data}, \citeyear{wing2019data} \cite{wing2019data}                & \cellcolor[HTML]{EFEFEF}\greencheckbox & \cellcolor[HTML]{EFEFEF}\greencheckbox & \cellcolor[HTML]{EFEFEF}\greencheckbox & \cellcolor[HTML]{EFEFEF}- & \cellcolor[HTML]{EFEFEF}- & \cellcolor[HTML]{EFEFEF}- & \cellcolor[HTML]{EFEFEF}\greencheckbox & \cellcolor[HTML]{EFEFEF}- & \cellcolor[HTML]{EFEFEF}\greencheckbox & \cellcolor[HTML]{EFEFEF}- & \cellcolor[HTML]{EFEFEF}- & \cellcolor[HTML]{EFEFEF}\greencheckbox & \cellcolor[HTML]{EFEFEF}\greencheckbox & \cellcolor[HTML]{EFEFEF}\greencheckbox \\
		
		& \citeauthor{ur2016big}, \citeyear{ur2016big} \cite{ur2016big}                & \redcheckbox & \redcheckbox & - & - & \redcheckbox & - & \redcheckbox & \redcheckbox & - & - & - & \redcheckbox & - & - \\
		
		& \cellcolor[HTML]{EFEFEF}\citeauthor{chen2014data}, \citeyear{chen2014data} \cite{chen2014data}                & \cellcolor[HTML]{EFEFEF}\bluecheckbox & \cellcolor[HTML]{EFEFEF}\bluecheckbox & \cellcolor[HTML]{EFEFEF}- & \cellcolor[HTML]{EFEFEF}- & \cellcolor[HTML]{EFEFEF}- & \cellcolor[HTML]{EFEFEF}- & \cellcolor[HTML]{EFEFEF}\bluecheckbox & \cellcolor[HTML]{EFEFEF}- & \cellcolor[HTML]{EFEFEF}\bluecheckbox   & \cellcolor[HTML]{EFEFEF}- & \cellcolor[HTML]{EFEFEF}- & \cellcolor[HTML]{EFEFEF}\bluecheckbox & \cellcolor[HTML]{EFEFEF}-   & \cellcolor[HTML]{EFEFEF}- \\
		
		& \citeauthor{jagadish2015big}, \citeyear{jagadish2015big} \cite{jagadish2015big}                & \bluecheckbox & \bluecheckbox & - & \bluecheckbox & \bluecheckbox & - & \bluecheckbox & - & \bluecheckbox   & - & - & \bluecheckbox & -   & \bluecheckbox \\
		
		& \cellcolor[HTML]{EFEFEF}\citeauthor{larson2016review}, \citeyear{larson2016review} \cite{larson2016review}                & \cellcolor[HTML]{EFEFEF}\bluecheckbox & \cellcolor[HTML]{EFEFEF}\bluecheckbox & \cellcolor[HTML]{EFEFEF}- & \cellcolor[HTML]{EFEFEF}- & \cellcolor[HTML]{EFEFEF}\bluecheckbox & \cellcolor[HTML]{EFEFEF}- & \cellcolor[HTML]{EFEFEF}\bluecheckbox & \cellcolor[HTML]{EFEFEF}\bluecheckbox & \cellcolor[HTML]{EFEFEF}-   & \cellcolor[HTML]{EFEFEF}\bluecheckbox & \cellcolor[HTML]{EFEFEF}- & \cellcolor[HTML]{EFEFEF}\bluecheckbox & \cellcolor[HTML]{EFEFEF}-   & \cellcolor[HTML]{EFEFEF}\bluecheckbox \\
		
		& \citeauthor{rizvi2017identifying}, \citeyear{rizvi2017identifying} \cite{rizvi2017identifying}                & \redcheckbox & \redcheckbox & \redcheckbox & \redcheckbox & \redcheckbox & - & - & \redcheckbox & -   & - & - & \redcheckbox & -   & - \\
		
		& \cellcolor[HTML]{EFEFEF}\citeauthor{demchenko2016cloud}, \citeyear{demchenko2016cloud} \cite{demchenko2016cloud}                & \cellcolor[HTML]{EFEFEF}\redcheckbox & \cellcolor[HTML]{EFEFEF}- & \cellcolor[HTML]{EFEFEF}\redcheckbox & \cellcolor[HTML]{EFEFEF}\redcheckbox & \cellcolor[HTML]{EFEFEF}\redcheckbox & \cellcolor[HTML]{EFEFEF}- & \cellcolor[HTML]{EFEFEF}\redcheckbox & \cellcolor[HTML]{EFEFEF}\redcheckbox & \cellcolor[HTML]{EFEFEF}-   & \cellcolor[HTML]{EFEFEF}\redcheckbox & \cellcolor[HTML]{EFEFEF}- & \cellcolor[HTML]{EFEFEF}\redcheckbox & \cellcolor[HTML]{EFEFEF}-   & \cellcolor[HTML]{EFEFEF}\redcheckbox \\
		
		& \citeauthor{wolf2016sage}, \citeyear{wolf2016sage} \cite{wolf2016sage}                      & \bluecheckbox & \bluecheckbox & \bluecheckbox & \bluecheckbox & -   & -   & \bluecheckbox & -   & -   & -   & -   & \bluecheckbox & -   & -  \\
		
		& \cellcolor[HTML]{EFEFEF}\citeauthor{sinaeepourfard2016towards}, \citeyear{sinaeepourfard2016towards} \cite{sinaeepourfard2016towards}                & \cellcolor[HTML]{EFEFEF}\greencheckbox & \cellcolor[HTML]{EFEFEF}\greencheckbox & \cellcolor[HTML]{EFEFEF}\greencheckbox & \cellcolor[HTML]{EFEFEF}\greencheckbox & \cellcolor[HTML]{EFEFEF}- & \cellcolor[HTML]{EFEFEF}- & \cellcolor[HTML]{EFEFEF}- & \cellcolor[HTML]{EFEFEF}- & \cellcolor[HTML]{EFEFEF}-   & \cellcolor[HTML]{EFEFEF}\greencheckbox & \cellcolor[HTML]{EFEFEF}- & \cellcolor[HTML]{EFEFEF}\greencheckbox & \cellcolor[HTML]{EFEFEF}\greencheckbox   & \cellcolor[HTML]{EFEFEF}- \\

		& \citeauthor{kim2016emerging}, \citeyear{kim2016emerging} \cite{kim2016emerging}                      & \bluecheckbox & \bluecheckbox & - & \bluecheckbox & \bluecheckbox   & -   & \bluecheckbox & \bluecheckbox   & \bluecheckbox   & -   & \bluecheckbox   & \bluecheckbox & -   & \bluecheckbox  \\

		& \cellcolor[HTML]{EFEFEF}\citeauthor{fisher2017selected}, \citeyear{fisher2017selected} \cite{fisher2017selected}                & \cellcolor[HTML]{EFEFEF}\bluecheckbox & \cellcolor[HTML]{EFEFEF}\bluecheckbox & \cellcolor[HTML]{EFEFEF}- & \cellcolor[HTML]{EFEFEF}- & \cellcolor[HTML]{EFEFEF}\bluecheckbox & \cellcolor[HTML]{EFEFEF}- & \cellcolor[HTML]{EFEFEF}\bluecheckbox & \cellcolor[HTML]{EFEFEF}- & \cellcolor[HTML]{EFEFEF}\bluecheckbox   & \cellcolor[HTML]{EFEFEF}\bluecheckbox & \cellcolor[HTML]{EFEFEF}\bluecheckbox & \cellcolor[HTML]{EFEFEF}\bluecheckbox & \cellcolor[HTML]{EFEFEF}\bluecheckbox   & \cellcolor[HTML]{EFEFEF}- \\

		& \citeauthor{turkay2018progressive}, \citeyear{turkay2018progressive} \cite{turkay2018progressive}                      & \bluecheckbox & \bluecheckbox & - & \bluecheckbox & \bluecheckbox   & -   & \bluecheckbox & -   & \bluecheckbox   & -   & -   & \bluecheckbox & -   & \bluecheckbox  \\

		& \cellcolor[HTML]{EFEFEF}\citeauthor{smith2017featurehub}, \citeyear{smith2017featurehub} \cite{smith2017featurehub}                & \cellcolor[HTML]{EFEFEF}- & \cellcolor[HTML]{EFEFEF}\yellowcheckbox & \cellcolor[HTML]{EFEFEF}\yellowcheckbox & \cellcolor[HTML]{EFEFEF}- & \cellcolor[HTML]{EFEFEF}- & \cellcolor[HTML]{EFEFEF}- & \cellcolor[HTML]{EFEFEF}- & \cellcolor[HTML]{EFEFEF}- & \cellcolor[HTML]{EFEFEF}-   & \cellcolor[HTML]{EFEFEF}\yellowcheckbox & \cellcolor[HTML]{EFEFEF}\yellowcheckbox & \cellcolor[HTML]{EFEFEF}\yellowcheckbox & \cellcolor[HTML]{EFEFEF}-   & \cellcolor[HTML]{EFEFEF}\yellowcheckbox \\

		& \citeauthor{wang2019human}, \citeyear{wang2019human} \cite{wang2019human}                      & \bluecheckbox & \bluecheckbox & - & \bluecheckbox & \bluecheckbox   & -   & \bluecheckbox & \bluecheckbox   & \bluecheckbox   & -   & -   & \bluecheckbox & -   & -  \\

		& \cellcolor[HTML]{EFEFEF}\citeauthor{lo2020systematic}, \citeyear{lo2020systematic} \cite{lo2020systematic}                & \cellcolor[HTML]{EFEFEF}\bluecheckbox & \cellcolor[HTML]{EFEFEF}\bluecheckbox & \cellcolor[HTML]{EFEFEF}- & \cellcolor[HTML]{EFEFEF}\bluecheckbox & \cellcolor[HTML]{EFEFEF}\bluecheckbox & \cellcolor[HTML]{EFEFEF}\bluecheckbox & \cellcolor[HTML]{EFEFEF}\bluecheckbox & \cellcolor[HTML]{EFEFEF}\bluecheckbox & \cellcolor[HTML]{EFEFEF}-   & \cellcolor[HTML]{EFEFEF}\bluecheckbox & \cellcolor[HTML]{EFEFEF}- & \cellcolor[HTML]{EFEFEF}\bluecheckbox & \cellcolor[HTML]{EFEFEF}-   & \cellcolor[HTML]{EFEFEF}- \\

		& \citeauthor{siva2020}, \citeyear{siva2020} \cite{siva2020}                      & \greencheckbox & \greencheckbox & - & \greencheckbox & \greencheckbox   & \greencheckbox   & \greencheckbox & \greencheckbox   & \greencheckbox   & \greencheckbox   & \greencheckbox   & \greencheckbox & -   & -  \\

		\multirow{-27}{*}{\rotatebox{90}{Team process}}                                  
		& \cellcolor[HTML]{EFEFEF}\citeauthor{stodden2020data}, \citeyear{stodden2020data} \cite{stodden2020data}                & \cellcolor[HTML]{EFEFEF}\greencheckbox & \cellcolor[HTML]{EFEFEF}\greencheckbox & \cellcolor[HTML]{EFEFEF}\greencheckbox & \cellcolor[HTML]{EFEFEF}\greencheckbox & \cellcolor[HTML]{EFEFEF}\greencheckbox & \cellcolor[HTML]{EFEFEF}\greencheckbox & \cellcolor[HTML]{EFEFEF}\greencheckbox & \cellcolor[HTML]{EFEFEF}- & \cellcolor[HTML]{EFEFEF}\greencheckbox   & \cellcolor[HTML]{EFEFEF}\greencheckbox & \cellcolor[HTML]{EFEFEF}\greencheckbox & \cellcolor[HTML]{EFEFEF}\greencheckbox & \cellcolor[HTML]{EFEFEF}\greencheckbox   & \cellcolor[HTML]{EFEFEF}\greencheckbox \\
		\hline
\end{tabular}
\end{table*}

\end{document}